\documentclass[11pt ]{JHEP3}

\usepackage{epsfig }

\title{Stringy NJL and Gross-Neveu models at finite density and temperature}
\author{Joshua L. Davis, Michael Gutperle, Per Kraus \\ Department of Physics and Astronomy, UCLA, Los Angeles, CA 90095, United States \\ E-mail:  \email{davis@physics.ucla.edu},\email{gutperle@physics.ucla.edu}, \email{pkraus@physics.ucla.edu} }
\author{Ivo Sachs \\ Arnold-Sommerfeld-Center for Theoretical Physics
Theresienstra§e 37, D-80333 M\"unchen, Germany \\ E-mail: \email{ivo@theorie.physik.uni-muenchen.de}}

\abstract{Nonlocal stringy versions of the Nambu-Jona-Lasinio and Gross-Neveu models
arise in a certain limit of holographic QCD.    We analyze the  phase structure at  finite density and temperature at strong coupling in terms of probe branes in the gravity dual.    Comparison
with the phase structure of the local field theory models shows qualitative agreement
with some aspects, and disagreement with others.   Finally, we explain how to construct the Landau potentials for these models by taking the probe branes off-shell. }

\preprint{  ASC-53/07}

\begin{document}

\section{Introduction}

Although most often studied as a candidate theory of quantum
gravity,  string theory has a rich history as a tool for
investigating the dynamics of strongly coupled quantum field
theories. The real-world application of most interest is QCD, an
infra-red confining gauge theory with a small number of colors and
flavors. Although the description of QCD in string theory remains
elusive, there has been  significant progress. A key
development was the realization by Maldacena
\cite{Maldacena:1997re} that the near-horizon dynamics of a large
number of $D3$-branes can be expressed in two different ways,
resulting in the famed duality between $\mathcal{N}=4$ super
Yang-Mills and IIB string theory on $AdS_5 \times S^5$. Since that
insight, systems of branes have been used to construct
gravitational duals to field theories with flavor \cite{Karch:2002sh}, with reduced
supersymmetry, and that exhibit confinement/deconfinement
transitions \cite{Witten:1998zw}. These developments have brought us closer to
capturing the dynamics of real-world QCD within string theory.

One natural target for such investigations is to use string theory
to learn about the phase diagram of QCD at finite temperature and
baryon number.  The QCD phase diagram has received much attention
in recent years, and a remarkably rich and intricate structure has
emerged, depending sensitively on the number of active flavors of
quarks in the theory; see \cite{RajagopalWF} for a review and
further references.  Possible applications include the physics of
heavy ion collisions and the structure of neutron stars. At the
same time, the theoretical tools for investigating these matters
are quite limited.  Lattice methods are powerful for studying
finite temperature QCD, but the notorious sign problem puts the
high density regime out of reach. At asymptotically high density
QCD becomes weakly coupled and tractable, but in the intermediate
density regime relevant to the real world there exist no
controlled calculational methods.

A popular avenue of attack is to use modern variants of the
Nambu-Jona-Lasinio model \cite{Nambu:1961tp} to study the
formation of various quark condensates (important in, for example, chiral symmetry breaking).  The main reason why NJL
models are relatively tractable is that they lack gluons and hence confinement.
While this obviously limits their application to many aspects of
QCD, at high density confinement does not play a dominant role
(\textit{e.g.} color can be spontaneously broken), and so NJL models are
useful phenomenological models (see \cite{Hatsuda:1994pi,Buballa:2003qv} for reviews of
the NJL model applied to QCD).

More recently, an embedding of NJL-like models in QCD has been suggested in  \cite{Antonyan:2006vw}.
By taking a decompactified limit of the model of Sakai and Sugimoto
\cite{Sakai:2005yt}, the authors of \cite{Antonyan:2006vw}
considered a D-brane system that cleanly exhibits chiral symmetry
breaking (see also \cite{Antonyan:2006qy,Antonyan:2006pg}). The brane construction of \cite{Antonyan:2006vw}
provides a gravitational dual to a theory we will refer to as the \textit{stringy NJL model}. Although the precise dynamics of this theory are unknown, the nomenclature follows from the appearance of chiral symmetry breaking and the lack of confinement.
%
The setup involves a large number of $D4$-branes which intersect a
stack of $D8$-branes and a stack of $\overline{D8}$-branes, these
stacks being separated in a direction along the $D4$-brane
worldvolume. In the regime in which the number of 4-branes (the
``color'' branes) is parametrically larger than the number of
8-branes (the ``flavor'' branes), the strong coupling dynamics of
the field theory can be easily studied using the DBI action of the
8-branes embedded in the near horizon background sourced by the
4-branes.\footnote{
A basic requirement for studying finite density QCD in string
theory is to include matter in the fundamental representation of
the gauge group.  For  $N_f/N_c \ll 1 $ this can be accomplished
by embedding probe branes in a supergravity background
\cite{Karch:2002sh}; for a sampling of further work in this
direction see
\cite{Babington:2003vm,Kruczenski:2003be,Sakai:2004cn}.} Unlike in the Sakai-Sugimoto model,  chiral symmetry breaking is not explicit in the decompactified theory but instead emerges dynamically as the recombination of the unstable
brane/anti-brane system, relating the separate $U(N_f)$ factors associated with each stack
of 8-branes.

The main purpose of the present work is to map out the phase
diagram of the stringy NJL model by considering the theory at
finite temperature and charge density. In the string description,
this is understood in terms of exciting the gauge fields that
propagate on the 8-brane worldvolume. This topic has been studied
in \cite{Horigome:2006xu,Parnachev:2006ev}\footnote{Recent paper addressing the same
question in the Sakai-Sugimoto model are \cite{Yamada:2007ys,Bergman:2007wp}.}, but we examine this
problem anew with an eye towards some important aspects not
considered previously. In particular, since the NJL model lacks
confinement there are finite energy massive quark states that are
expected to play a role in the phase diagram.  To incorporate
this, we find a new brane solution that describes a condensate of
smeared fundamental strings stretching from the 8-branes to the
horizon of the 4-brane geometry. Additionally, we study the
possibility of non-homogeneous phases in which the field theory is
in a configuration where regions of charged vacuum and uncharged
vacuum coexist. It turns out that such mixed phases dominate the
thermodynamics.  We map out the resulting phase diagram and
compare with an analogous phase diagram of a field theory NJL
model.

There are of course many limitations and caveats in attempting to
apply this approach to real QCD.   Among these, we need to be able
to move away from the $N_c \gg 1$,  $N_f\ll N_c$ regime, to
incorporate bare quark masses, and to better understand the
extrapolation from strong to weak coupling. These are all serious
obstacles, but given the richness of finite density QCD combined
with the scarcity of reliable methods for its study, we feel that
there is more insight to be gained from developing the stringy
models considered here.

While the relation between the D-brane constructions discussed here and the local NJL model is quite
subtle (see \cite{Antonyan:2006pg} for discussion), the situation is much clearer for its $1+1$ dimensional cousin, the Gross-Neveu (GN) model \cite{Gross:1974jv}. The stringy version of the GN-model was  constructed in
\cite{Antonyan:2006qy} as intersecting D4 and D6 branes.
The study of this theory at finite charge density is
especially simple.  In field theory, bosonization reduces the
problem to one involving a free scalar field.  We derive the
analogous result in the string version, which involves the
dynamics of pure gauge degrees of freedom.  The massless Goldstone excitation which relates to the phase of the chiral condensate in the field theory description is mapped to a pure gauge potential on the probe D6 brane whereas the massive fermions correspond to gauge potentials that are sourced by fundamental strings ending on the probe branes. As a consequence of the Chern-Simons term on the 8-brane they do not, however, induce charges on the non-local GN-model at the boundary of the AdS-space.

Finally we consider the problem of constructing off-shell Landau potentials for the the order parameter in these models. We  show that the (necessarily not unique) Landau effective potential for the order parameter can be constructed by taking the  probe-brane profile off-shell. In the GN model this potential can be used to reproduce the ``old fashioned" phase diagram of the GN-model in the mean field approximation while in the D4/D8- brane model it can be used to study the phase structure at zero temperature.

This paper is organized as follows. In section 2 we briefly review the $D$-brane construction of \cite{Antonyan:2006vw} leading to the stringy NJL model. In section 3 we consider the strongly coupled regime of this model, described by supergravity. We find solutions of the probe 8-branes with non-trivial worldvolume gauge fields. We compare the actions of these solutions in section 4, and so map out the phase diagram of the model. In section 5 we include a short but thorough review of the thermodynamics of the NJL field theory, deriving its phase diagram. We compare the phase diagrams of the field theory and supergravity descriptions in section 6. In section 7 we comment on the role of the 6-brane gauge fields in the stringy Gross-Neveu model. The construction and analysis of off-shell Landau potentials for the stringy NJL and GN models are discussed in section 8. We  close with some comments in section 9.

\section{Weakly coupled $D$-brane description}

Here we briefly review the NJL  $D$-brane  construction of
\cite{Antonyan:2006vw} (for the stringy Gross-Neveu model see
\cite{Antonyan:2006qy}). In the limit of weak string coupling, the
system is composed of three separate stacks of $D$-branes in
$\mathbb{R}^{1,9}$. There are $N_c$ coincident $D4$-branes
extended in the $\{x_0,x_1,x_2,x_3,x_4\}$ directions.\footnote{In
the Sakai-Sugimoto model \cite{Sakai:2005yt} the $x_4$ direction
is compactified. The present configuration can be considered a
limit of that model in which the confinement scale of the dual
field theory is taken to zero.} Furthermore, there are two stacks of
coincident $D8$-branes and coincident $\overline{D8}$-branes, each
with $N_f$ branes. The 8-branes are extended along all directions
but $x_4$; in the $x_4$ direction the stacks are separated by some
distance $L$ which we take to be much larger than the string
length $\ell_s$. We note that this configuration preserves no
supersymmetry.

We will study this system in the limits  of low energy and weak
string coupling. We consider a finite number $N_f$ of flavor
8-branes and an infinite number of color branes, taking the 't
Hooft limit of the 4-brane system, \textit{i.e.} $g_s \to 0$ and
$N_c\to \infty$ with $g_s N_c$ held fixed and finite. In this
regime, the 't Hooft coupling on the 8-branes vanishes, and so the
$U(N_f)_L \times U(N_f)_R$ gauge symmetry becomes an effectively
global symmetry.

There are two $(3+1)$-dimensional intersections of the 8-branes with the $D4$-branes. At low energy, the spectrum of open 4-8 strings consists only of massless chiral fermions localized at each intersection, left-handed for the $D8$-branes and right-handed for the $\overline{D8}$-branes. All the fermions are in the fundamental representation of $U(N_c)$, while the left-handed (right-handed) fermions are in the fundamental of the $U(N_f)_L$ ($U(N_f)_R$) factor of the global flavor symmetry and are uncharged under the other factor.

Fermions of opposite chirality interact  via exchange of the
$(4+1)$-dimensional $U(N_c)$ gluons of the $D4$-brane worldvolume
across the gap between the stacks of 8-branes. In the limit of
weak 't Hooft coupling ($g_s N_c  \ll 1$) we imagine integrating
out the $(4+1)$-dimensional gluons to yield a theory of pure
quarks where the $U(N_c)$ symmetry is now understood to be global.
The precise dynamics in this limit is not known and we will use a
standard local NJL quantum field theory as a toy model. In the
limit of large 't Hooft coupling, the dynamics are best captured
by supergravity. Specifically, the 4-branes source a gravitational
and 4-form magnetic flux background. The low-energy physics is
described by the DBI action of the 8-branes in the near-horizon
region of this background, which we discuss presently.

\section{Strong coupling analysis of the stringy Nambu-Jona-Lasino model}
The strong coupling description of the finite-temperature stringy NJL model with $N_c \gg N_f$ is encoded in the classical dynamics of $N_f$ probe 8-branes in the near horizon geometry of $N_c$ non-extremal 4-branes. The worldvolume of the
4-branes lies along the Euclidean time $\tau$ and the $x_i$ directions, where $i=1\ldots4$. The geometry and dilaton are given by\footnote{We use units in which $\alpha^\prime=1$.}
\begin{eqnarray}\label{metricd4}
ds^{2}&=& \left({U\over R}\right)^{3\over2} \Big( f(U) (d\tau )^{2} + (dx_{i})^{2}\Big)  +\left({U\over R}\right)^{-{3\over2}}\Big( {dU^{2} \over f(U)}+ U^{2}ds^2_{S_{4} } \Big)~, \nonumber  \\
e^{\phi}&=& g_{s}  \left({U\over R}\right)^{3\over4}~,
\end{eqnarray}
where the non-extremality of the 4-brane background is encoded in the function
\begin{equation}\label{fufun}
f(U)= 1-{U_{T}^{3 }\over U^{3}}~.
\end{equation}

The horizon of the black brane is located at $U=U_T$. The
absence of a conical singularity at the horizon determines the
periodicity of Euclidean time $\tau\sim \tau +\beta$, and hence the temperature, with
\begin{equation}\label{betadef}
\beta = {1\over T} ={4\pi R^{3\over 2} \over 3 U_{T}^{1/2}}~.
\end{equation}
The limit of zero temperature corresponds to $U_{T}=0$ and hence $f(U)=1$.

\FIGURE[r]{\epsfig{file=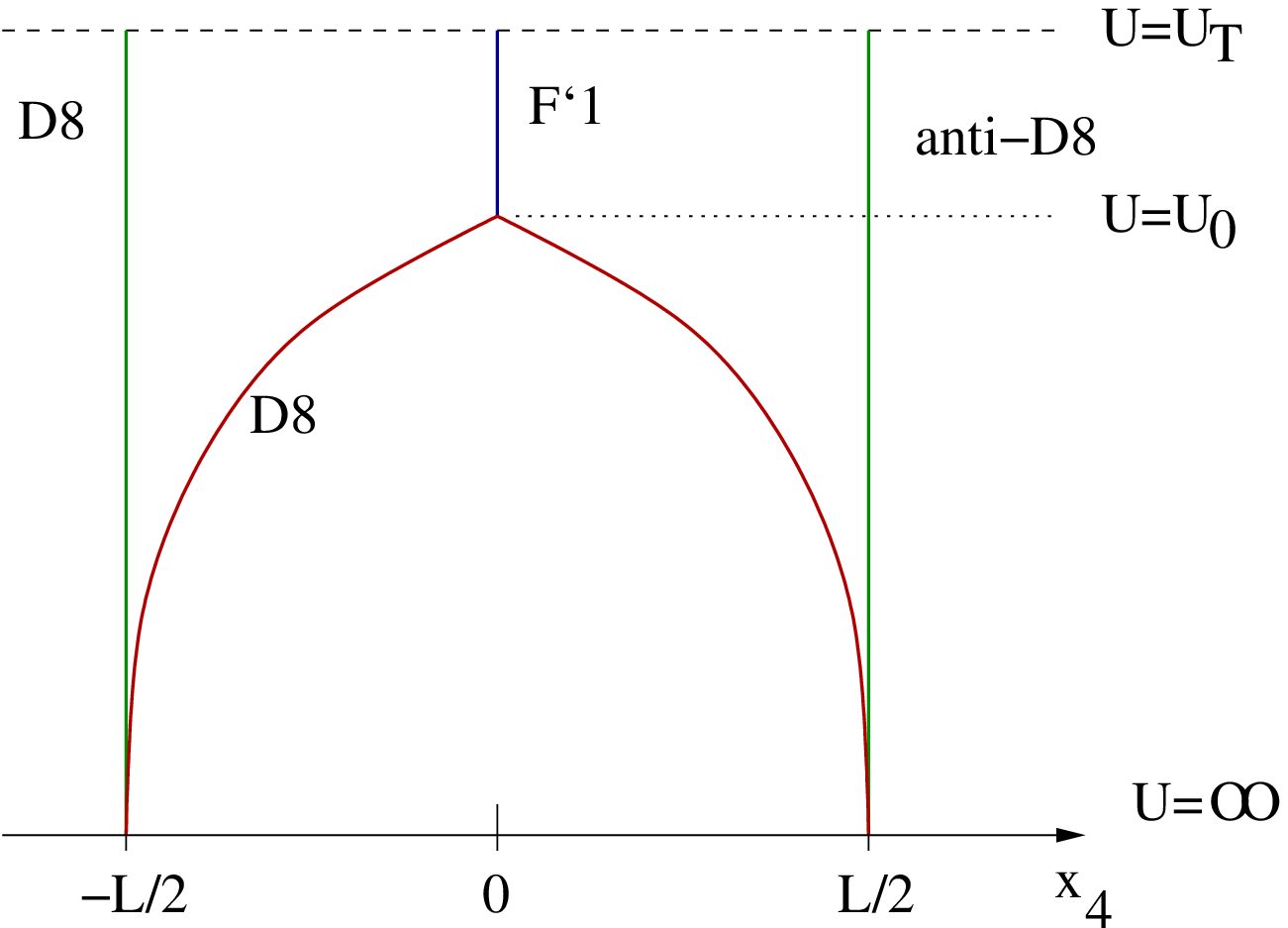, width=7cm}
        \caption{Probe $D8$ branes in the $D4$ brane background with attached F-strings.}\label{branepicture}}

As discussed in the previous section, chiral and anti chiral fermions are introduced by placing
flavor $D8$ and $\overline{D8}$-branes in the $D4$-brane background
\cite{Sakai:2004cn,Sakai:2005yt}. There are two qualitatively distinct configurations of D8 branes.
First, there are straight $D8$ and $\overline{D8}$-branes separated in $x_4$
direction, which extend all the way to the horizon at $U=U_T$. For
the straight branes the $U(N_f)\times U(N_f)$ chiral symmetry is
unbroken. Second, the $D8$-brane and anti-brane stacks can recombine and
produce a single curved $D8$-brane stack which does not reach the horizon. For
the curved brane the chiral symmetry is broken to $U(N_f)$
\cite{Antonyan:2006vw}.

A finite chemical potential for fermions can be introduced by an
imaginary gauge field $A_0= i \mu$ on the worldvolume of the 8-branes \cite{Parnachev:2006ev}.  We will discuss the straight and curved brane solutions with chemical potential in the following. One important observation
is that it is necessary to introduce fundamental strings ending on the 8-brane to obtain non-zero charge density in the chirally broken phase.

\subsection{Straight brane}

The straight (anti)-$D8$ branes  have a fixed $x_4$ position and lie along the $\tau,x_1,x_2,x_3$ as well as the $U$ and the $S^4$ directions, which we use to parameterize the worldvolume. We choose the $D8$-brane stack to be localized at $x_4=-L/2$ and the anti-$D8$-brane stack at $x_4=L/2$. The worldvolume of the branes reaches the horizon at $U=U_T$. To include a chemical potential in the dual description, we turn on the 8-brane gauge field $A_0$ and assume all other components vanish. The Dirac-Born-Infeld action for the $D8$-brane as well as the $\overline{D8}$-branes is then given by
\begin{equation}\label{bistraight}
S_s= \kappa  R^{3\over 2} \int_{U_T}^\infty  dU  U^{5\over 2} \sqrt{1+ \left( A'_0 \right)^2}~,
\end{equation}
where primes denotes derivatives with respect to $U$. The constant coefficient $\kappa$ is given by
\begin{equation}
\kappa = {T_8 \over g_s} N_f \beta V_3 \Omega_4 ~,
\end{equation}
where $T_8$ is the 8-brane tension and $V_3$ and $\Omega_4$ are the volumes of the three noncompact directions $\{x_{1,2,3}\}$ and the unit $S^4$, respectively.

Varying the action (\ref{bistraight}), we find an integral of motion (constant in $U$)
\begin{equation}\label{qstraight}
i Q_s = {\delta S\over \delta A'_0 }=\kappa  R^{3\over 2} {  U^{5\over 2} A'_0 \over \sqrt{1+ \left( A'_0 \right)^2}}~.
\end{equation}
We will call $Q_s$ the ``conserved charge'', although it is more properly
the electric flux on the brane. We can integrate (\ref{qstraight}) to solve for the gauge field. This relates the conserved charge and chemical potential $\mu=iA_0(U=\infty)$ via
\begin{equation}\label{mustraight}
\mu =Q_s\int_{U_T}^\infty \! dU~ {1
 \over\sqrt{  \kappa^2 R^3 U^5 + Q_s ^2  }}~,
\end{equation}
where we have set $A_0(U_T)=0$ to avoid a singularity at the horizon.

The combined action of the branes and anti-branes in terms of the conserved charge is given by
\begin{equation}\label{sstraight}
S_s = 2 \kappa R^{3/2} \int_{U_T}^\infty\! dU {U^5 \over
\sqrt{ U^5 +{Q_s^2\over \kappa^2 R^3}} }~.
\end{equation}
Note that both the brane separation $L$ as well as  the non-extremality function $f(U)$ are irrelevant for the straight brane action. For the straight brane as well as the curved brane, the Chern-Simons term $\int A\wedge dA \wedge F_6$ on the $D8$-brane worldvolume can be neglected, since the only nontrivial gauge field component is $A_0$.

\subsection{Curved brane}

The brane configuration corresponding to broken chiral symmetry is a curved $D8$-brane. We treat this similarly to the straight brane configuration except that we use the coordinate $x_4$ instead of $U$ to parameterize the worldvolume. The embedding is determined by the profile function $U(x_4)$ with boundary conditions such that $U(x_4 \to \pm L/2)\to \infty$. The curved brane is symmetric about $x_4=0$, where its position $U(x_4=0)\equiv U_0$ is closest to the horizon.
As we shall see it is necessary to introduce fundamental string ending on the $D8$-brane to achieve a nonzero charge density.

The Dirac-Born-Infeld action for the $D8$-brane in this parametrization is
\begin{equation}\label{curveds}
S_c=\kappa \int_{-L/2}^{L/2}\; dx_4 ~ U^4 \sqrt{f(U)+ \left({U \over
R}\right)^{-3}(\dot{U}^2+\dot{A}_0^2)}~,
\end{equation}
where the dot denotes a derivative with respect to $x_4$. There are integrals of motion independent of $x_4$
\begin{eqnarray}\label{conservc}
iQ_c &=& {\delta S_c \over \delta  \dot{A}_0}=\kappa R^3 { U \dot{A}_0 \over \sqrt{f(U)+ \left({U
\over R}\right)^{-3}(\dot{U}^2+\dot{A}_0^2)}}~, \nonumber \\
P_4 & =&  L-  {\delta S_c \over \delta  \dot{U}}  \dot U  -{\delta S_c \over \delta  \dot{A}_0}  \dot{A}_0= \kappa {
U^4 f(U) \over \sqrt{f(U) + \left({U \over
R}\right)^{-3}(\dot{U}^2+\dot{A}_0^2)}}~.
\end{eqnarray}
Once again we refer to these integrals of motion as conserved charges. In order to obtain the same chemical potential on both branches of the D8 brane  the gauge field $A_0$ must be  symmetric about $x_4=0$. This implies that, for a smooth gauge field, $\dot A_0=0$ at $x_4=0$. It follows from (\ref{conservc}) that $Q_c=0$ and hence $A_0$ has to be constant. Since the Chern-Simons term is not relevant for the stringy NJL model (in contrast to the stringy GN model, as we will see later) the constant gauge field can be set to zero and hence it seems that the chemical potential has a trivial effect on the dynamics.

A way to overcome this is to loosen the condition that the gauge field $A_0$ and the brane embedding $U(x_4)$ must be smooth at $x_4=0$. This can be achieved by introducing fundamental strings located at $x_4=0$ stretching from the horizon to the worldvolume of the $D8$-brane resulting in the configuration shown in Figure \ref{branepicture}.
The action for $N_1$ fundamental strings is (recall we set $\alpha^\prime=1$)
\begin{eqnarray}\label{funds}
S_F &=& {N_1  \over 2\pi } \int\! \sqrt{\det h}
+{iN_1 \over 2\pi } \int\! d\tau  A_0~, \nonumber \\
 &=& {N_1 \beta  \over
2\pi } \int\!  dx_4~ [U(x_4)-U_T] \delta(x_4)  +{iN_1 \beta
\over 2\pi } \int\!  dx_4~ \delta(x_4 )  A_0~.
\end{eqnarray}
The variation of the combined action of fundamental strings and $D8$-branes leads to the jump conditions for the $Q_c$ and $\dot U$  at $x_4=0$
\begin{eqnarray}\label{jumpa}
\Delta  Q_c & =& 2 Q_c= {N_1 \beta \over 2\pi }~, \nonumber \\
 \Delta \dot{U} &=& 2 \dot U\mid_{x_4=0}=  { f(U_0)\over P_4 } \left({U_0\over R}\right)^{3} {N_1 \beta \over 2\pi }~,
\end{eqnarray}
where we used the fact that for the symmetric $D8$-brane the gauge field derivative $\dot A_0$ and $\dot U$ change sign at $x_{4}=0$. We choose the charge on the right side of the $D8$-brane (where $x_{4}>0$) to be positive. Using (\ref{jumpa}), together with (\ref{conservc}), one can show that $\dot U =-i \dot A_0 $ at $x_4=0$. Substituting this back into the expression for $P_4$ given in (\ref{conservc}) one finds
\begin{equation}\label{jumpb}
  P_4  = \kappa U_0^4 \sqrt{f(U_0)}~.
\end{equation}
The expression of the conserved charges (\ref{conservc}) can be solved for the derivatives $\dot A_{0}$ and $\dot{U}_{0}$
\begin{eqnarray}\label{jumpc}
 \dot{U} & = &{f(U) \over \kappa U_0^4 \sqrt{f(U_0)}}\left({U\over
R}\right)^{3\over 2}\sqrt{\kappa^2 U^8+\left({U\over R}\right)^{3}
Q_c^2-{\kappa^{2 }U_{0}^{8} f(U_{0}) \over f(U)}}~, \nonumber \\
-i\dot{A}_0 &=&  {Q_c\over \kappa} {f(U)\over \sqrt{f(U_0)} U_0^4} \left({U\over R}\right)^3~.
\end{eqnarray}
Integration of  the first equation in  (\ref{jumpc}) relates the conserved charges to the width separation $L$ in the $x_{4}$ direction of the two asymptotic legs of the $D8$-brane.
\begin{eqnarray}\label{conca}
 L &=& 2 \int_{U_0}^\infty { dU \over \dot{U}}~, \nonumber \\
&=&  2  U_0^4 \sqrt{f(U_0)} \int_{U_0}^\infty  {dU \over   f(U) \left({U\over
R}\right)^{3\over 2}\sqrt{  U^8+\left({U\over R}\right)^{3}
{Q_c^2\over \kappa^2}-{  U_0^8 {f(U_0)} \over f(U)}}}~.
\end{eqnarray}
Integration of  the second equation in (\ref{jumpc}) relates the conserved charges and the chemical potential
\begin{eqnarray}\label{concb}
\mu& =& -i A_{0}(U_{0})+\int_{U_0}^\infty    dU  \; {-i \dot A_0 \over \dot{U}}~, \nonumber \\
&=&    -i A_{0}(U_{0})+{Q_c\over \kappa}
\int_{U_0}^\infty {dU \over \left({U\over R}\right)^{-{3\over
2}}\sqrt{  U^8+\left({U\over R}\right)^{3} {Q_c^2\over \kappa^2}-{  U_0^8 {f(U_0)} \over
f(U)}}}~.
\end{eqnarray}
These two equations relate the parameters governing the curved brane: $L, \mu , Q_c, U_0$ and $T$. For example at fixed temperature, (\ref{conca}) and (\ref{concb}) can be used to find the turning point $U_0$ and the charge $Q_c$, given the separation $L$ and the chemical potential $\mu$ and vice versa.

\section{Phase diagrams at strong coupling}

The solutions discussed in the last section provide strong coupling descriptions of possible phases for the stringy NJL model. To determine which phase dominates for specific external conditions, we must compare their free energies (or other appropriate thermodynamic potentials). In the language of the previous section, this entails comparison of on-shell actions (or appropriate Legendre transforms).

When including the effects of non-zero gauge field on the probe 8-branes, one can perform the comparisons at fixed charge density or fixed chemical potential. We will analyze the two cases in turn.

\subsection{Fixed chemical potential}
\FIGURE[r]{\epsfig{file=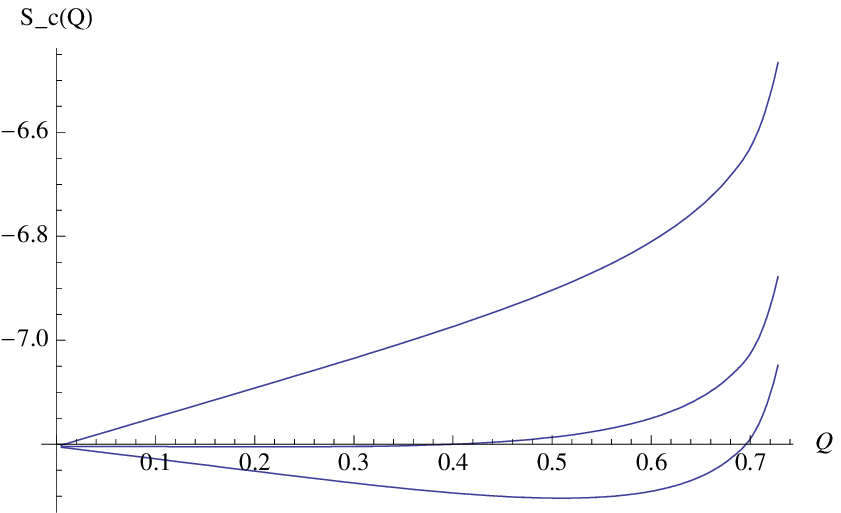, width=7cm}\caption{ Action of the curved brane  as a function of the charge $Q$ at $T=.3$ and $L=.5$ for various values of the chemical potential $\mu$}. \label{figmudep}}

The thermodynamic potential which governs the phase structure at fixed chemical potential is $\Omega(T,\mu) =E-TS-\mu Q$, the \textit{grand free energy}. In the semiclassical probe approximation the grand free energy of the system at fixed chemical potential is given by the Dirac-Born-Infeld action of the branes and the Nambu-Goto action of the fundamental string, where the boundary value of the imaginary gauge field $A_{0}(U=\infty)=i\mu$ determines the chemical potential. The actions are formally divergent so in order to evaluate them we introduce an ultraviolet cutoff at $U=\Lambda$. All actions have the same divergence in $\Lambda$ and so the difference of the grand free energies is independent of the cutoff as $\Lambda\to \infty$.

In the following we choose units of length and action such that $R=1$ and $\kappa=1$.
The grand free energy for the unbroken phase is given  by the action  of the straight branes (\ref{sstraight})  and reads
\begin{equation}\label{sstraib}
S_{s}(\mu)= 2 \Big(\int_{U_{T}}^{\Lambda} dz {z^{5}\over \sqrt{z^{5}+ Q_{s}^{2}}}-{2\over 7}\Lambda^{7/2}\Big)~.
\end{equation}
The charge density $Q_{s}$ on the straight branes is determined uniquely by the chemical potential through (\ref{mustraight}).

The grand free energy for the broken phase is given by the sum of the action of the curved brane (\ref{curveds}) and the smeared fundamental strings (\ref{funds})
\begin{eqnarray}\label{scurvb}
S_c(\mu, Q_c)&=& 2\Big(\int_{U_0}^{\Lambda} dz {z^5 \over \sqrt{z^5 -{U_0^8 f(U_0)\over z^3 f(z) }+Q_{c}^2} }
+ Q_{c}(U_0-U_T)+ i A_{0}(U_{0}) Q_{c}- {2\over 7}\Lambda^{7/2}\Big)~.\nonumber\\
 \end{eqnarray}
Unlike the straight brane scenario, the charge density is not determined uniquely in terms of the boundary data (the chemical potential $\mu$ and the asymptotic brane separation $L$). Rather, even after imposing the constraints (\ref{conca}) and (\ref{concb}) to eliminate the location $U_0$ of the tip of the curved brane and the value of the gauge field at the tip $A_0(U_0)$, the charge is still free to be varied. The charge on the curved brane is then that which minimizes the (\ref{scurvb}) for a fixed chemical potential. The behavior of the action displayed in figure \ref{figmudep} is generic; for small chemical potential the $Q_c=0$ curved brane is dominant whereas for larger chemical potential the curved branes with $Q_c\neq 0$ are dominant.

\FIGURE[r]{\epsfig{file=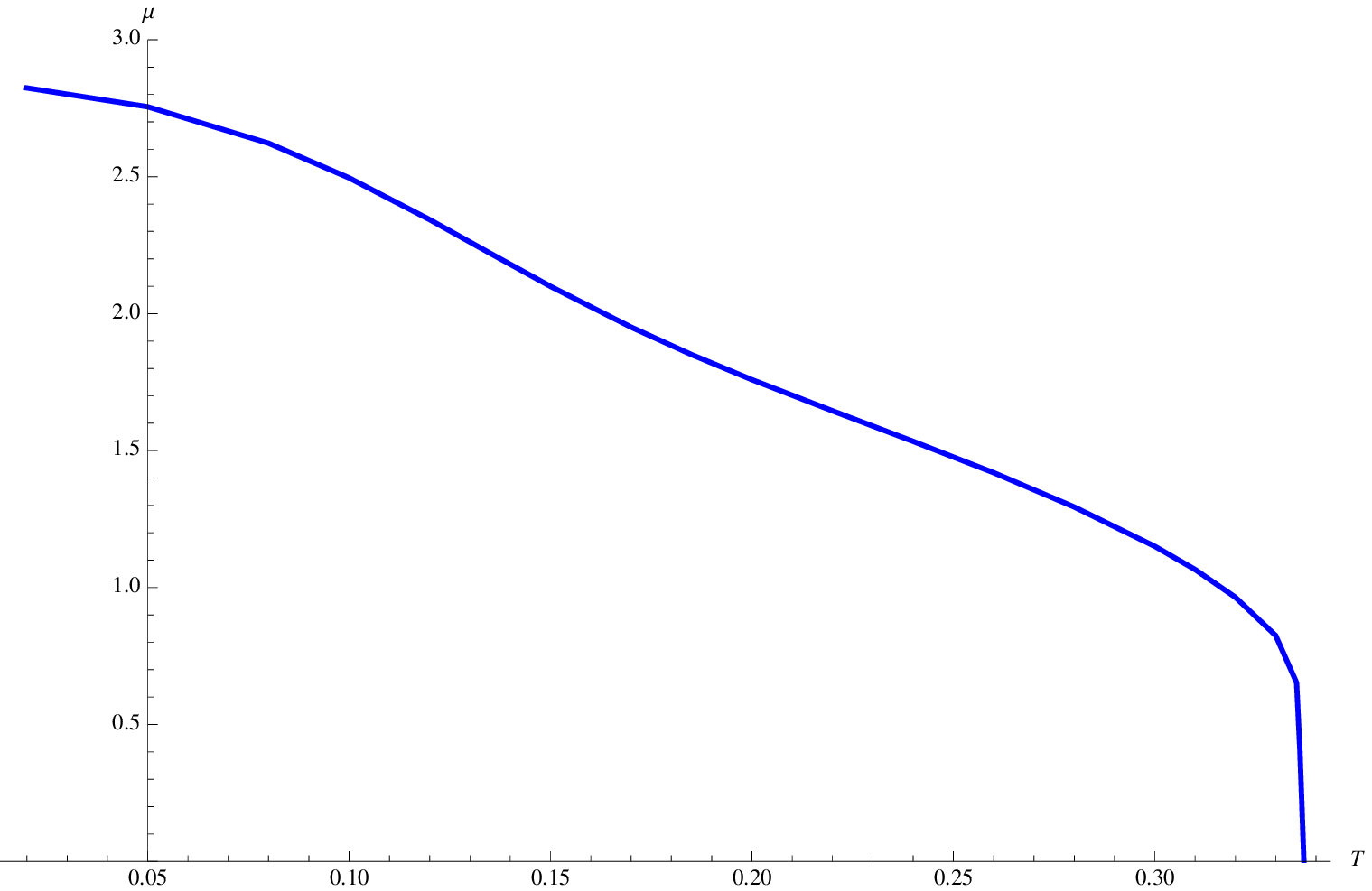, width=7cm}
        \caption{Phase diagram at fixed chemical potential. Below the plotted curve, chiral symmetry is broken. Above the curve, the symmetry is restored.}\label{plotphmu}}

To obtain the phase diagram of the system, we rely on numerical comparison of the actions discussed above. We find that at sufficiently low temperature, the curved brane without charge density is thermodynamically preferred and so chiral symmetry is broken. Raising the temperature and/or potential, one encounters a transition to the straight brane (\textit{i.e.} chiral symmetry is restored), whose charge is determined uniquely by the chemical potential. Although the $Q_c\ne0$ phase of the curved brane does eventually become preferred over the $Q_c=0$ phase, it appears that it never is preferred over the straight brane phase. We show the phase diagram for the system in figure \ref{plotphmu}.

The phase boundary is given by the curve in $\mu,T$ space where $S_{c}(\mu , Q_c=0)=S_{s}(\mu)$. The phase diagram displayed in figure \ref{plotphmu} shows this phase boundary, the critical value of the chemical potential $\mu_{phase}$ as a function of the temperature $T$. At a critical temperature of $T=.336$ the curved brane ceases to exist and the system is in the unbroken phase. The plot is presented for a specific value of the separation $L$. However as we shall argue in the end of this section, a scaling argument shows that the phase diagram is universal.

\subsection{Fixed charge}
In the previous section we worked at fixed chemical  potential,
where  the  thermodynamic potential which governs the phase
structure  is $\Omega(T,\mu) =E-TS-\mu Q$. If instead we work at
fixed charge the new thermodynamic potential is the Helmholtz free energy, given by a
Legendre transform $F(T,Q)= \Omega+ \mu Q$.  On the
gravity side this is accomplished by adding $\mu Q= -i
A_{0}(U=\infty) Q$ to the action. Note that for a given fixed charge, the
gauge field $A_{0}(U=\infty)$ and hence the chemical potential $\mu$ is not the same for
the straight and curved brane, but is determined in terms of the
charge by the constraints (\ref{mustraight}) and (\ref{concb})
respectively.

The modified action for the straight brane is given by
\begin{equation}\label{sstrc}
\tilde S_{s}(Q)= S_{s}+ \mu_{s} Q =  2 \Big( \int_{U_T}^\Lambda \! dU
\sqrt{ U^5 +Q^{2}}- {2\over 7}\Lambda^{7/2}\Big)~.
\end{equation}
The modified action for the curved brane with fundamental strings attached
\begin{eqnarray}\label{curvacb}
\tilde S_c(Q)&=&S_{c}+ \mu_{c} Q=2\Big(\int_{U_0}^{\Lambda} dz {z^5 + Q^2 \over \sqrt{z^5 -{U_0^8 f(U_0)\over z^3 f(z) }+Q^2} }+ Q(U_0-U_T)- {2\over 7}\Lambda^{7/2}\Big)~.
\end{eqnarray}
As before the phase diagram can be obtained  by considering the
sign of $\Delta \tilde S= \tilde S_{s}-\tilde S_{c}$. The  phase
boundary in the $Q,T$ plane for separation $L=0.5$ is plotted in
figure \ref{figurefour}. It is an interesting feature of the phase
diagram that the critical value of the charge  increases as
the temperature is increases.

\FIGURE[r]{\epsfig{file=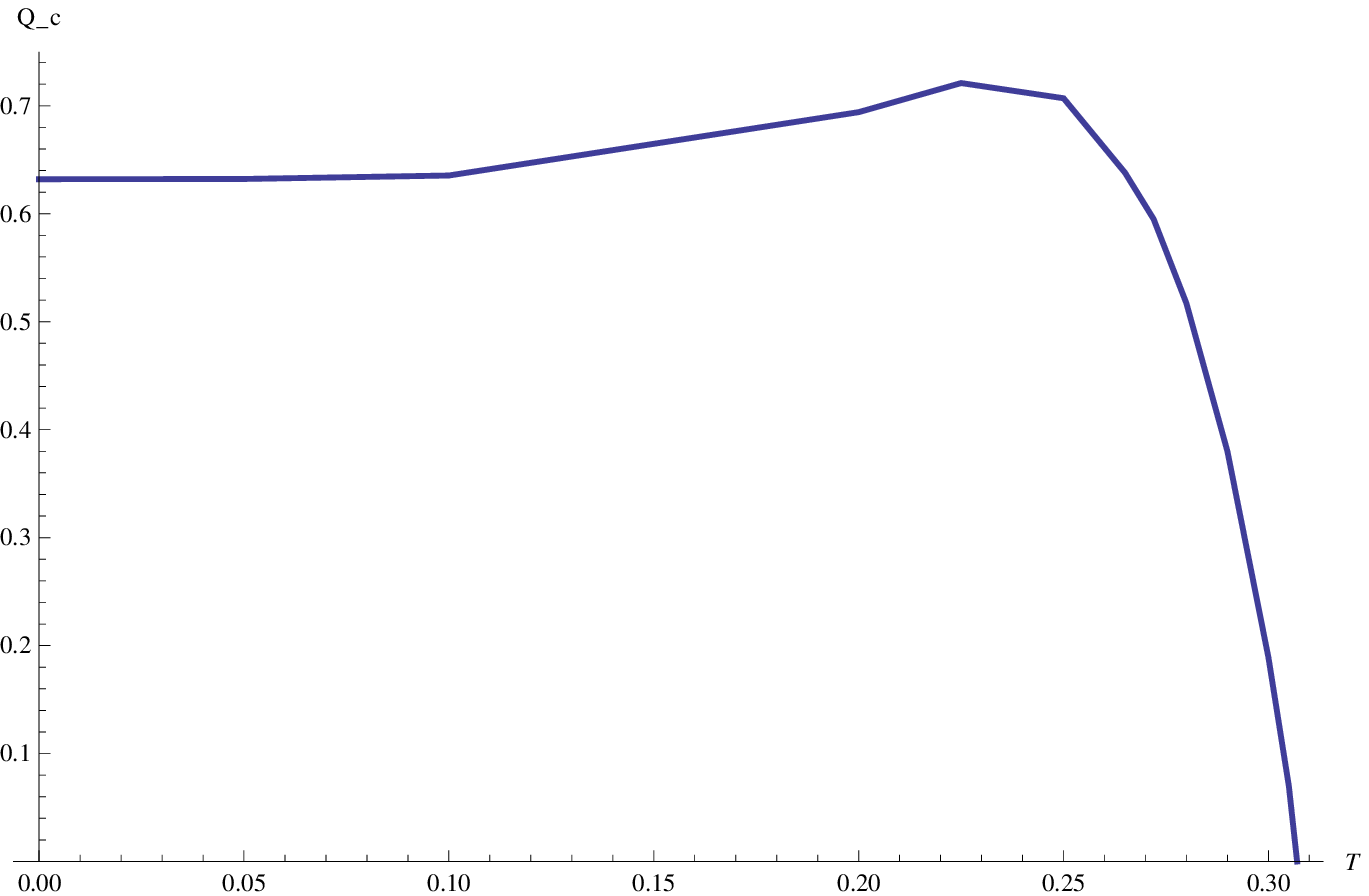 , width=7cm}
        \caption{Phase diagram for fixed charge  with $L=0.5$.}\label{figurefour}}

The numerical  analysis of the phase structure in the last two
sections  was done at a fixed value  $L=0.5$ of the separation.
The phase structure is however universal in the following sense.
If we scale $L\to L \alpha$ along with
\begin{equation} \label{scalea}
U \to U/\alpha^2, \quad Q\to Q / \alpha^5, \quad \mu\to \mu /\alpha^{2}~,
\end{equation}
the constraints (\ref{mustraight}), (\ref{conca}) and (\ref{concb}) are invariant. Under these scalings the actions (\ref{sstrc}) and (\ref{curvacb}) both scale as $S \to S / \alpha^7$.
Since the phase structure is only sensitive to the relative sign
of the difference of actions the phase diagram remains the same
under the scaling. Although the solutions depend in general on
three parameters $L,T,\mu$ at fixed chemical potential and $L,T,
Q$ at fixed charge, the scaling symmetry can be used to fix one of
them, which we have done.

\subsection{Mixed phase}
In the case of fixed charges we considered two phases: the straight branes and curved brane with constant charge density. As discussed earlier the curved brane must have fundamental strings ending on it, in order to support a nonzero charge density.

There is however the possibility to have a inhomogeneous mixed phase with regions of broken chiral symmetry and regions where the symmetry is restored. The charge is carried by the straight brane regions and the
curved brane regions do not have any fundamental strings ending on
them.  If the total charge density is $Q$ and the straight brane
takes up $y V$ of the volume $V$ whereas the curved brane with no
charge takes up  $(1-y) V$ of the volume, the charge density on
the straight brane has to be  $Q/y$. If the contribution of the
interface energy is negligible (which seems sensible in the large
volume limit since the bulk contribution scales as the volume
whereas the interface scales as the area) the free energy in the
mixed phase is given by

\begin{eqnarray}
S_m(Q,y)&=& 2\Big( y \int_{U_T}^\Lambda dz\sqrt{z^5+\left({Q\over y}\right)^2}+ (1-y) \int_{\bar U_0}^\Lambda dz {z^5\over \sqrt{ z^5 - {\bar U_0^8 f(U_0)\over z^3 f(z)}}}- {2\over 7}\Lambda^{7/2}\Big)~.\nonumber \\  \label{mixedaction}
\end{eqnarray}

The free energy of the mixed phase should be compared to the free
energy of the curved and straight brane at the same total charge
density. Numerically we have investigated this question and we
checked that for all our choices of charges and temperatures  the
mixed phase has smaller free energy than the curved brane with the
same charge.
\FIGURE[r]{\epsfig{file=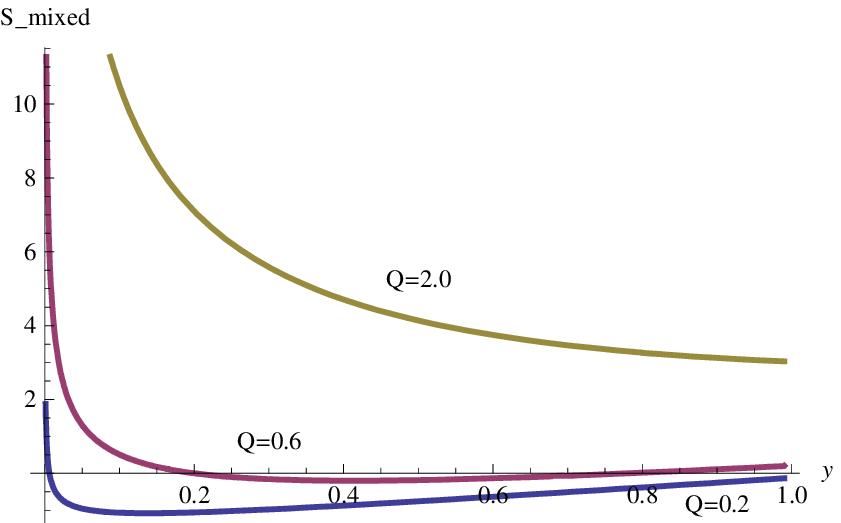 , width=7cm}
        \caption{Free energy versus fraction $y$ of system in chirally symmetric phase. Plots for $L=0.5,T=0.2$ and three distinct charges $Q=0.2,Q=0.6,Q=2.0$.}\label{figurefive}}
Hence  for parameters where the previous analysis
showed that the chiral symmetry is broken, the preferred phase of
the system is mixed. Incidentally  for some range of charges this is also true
for the phase where the straight brane dominates since the free
energy of the straight brane is given by setting $y=1$ for the
mixed phase. For small charges even then the mixed phase dominates. If the charge is increased the straight brane will have smaller free energy and dominate.   A representative plot of the free energy as a function of $y$  for various values of the charges is
given in figure \ref{figurefive}. Note that at large enough charge the minimum free energy is at $y=1$, indicating a homogeneous phase of restored chiral symmetry.

\section{NJL model at finite temperature and chemical potential}

In the previous section we studied the finite temperature/chemical
potential phase diagram of the stringy NJL model at strong
coupling.  Now we want to compare against an analogous phase
diagram in ordinary field theory.   The field theory in question
will be the usual NJL model at large $N$.   Before turning to the
calculations we pause to motivate our study of the NJL model in
the present context.  As noted in the introduction, the eventual
goal is to learn something about finite density QCD.  The phase
structure of high density QCD is for the most part governed by the
condensation of various fermion bilinears.  For instance, at
sufficiently high density it is known that a diquark condensate
forms, spontaneously breaking the gauge symmetry (``color
superconductivity").  The formation of such condensates can be
studied in the context of NJL models with four-fermion
interactions.  Indeed, at asymptotically high density such a
treatment is under good control due to asymptotic freedom.  At the
smaller densities relevant to the real world, controlled
computations are harder to come by, but NJL models still serve as
valuable phenomenological guides, although one should be aware of
ambiguities due to cutoff dependence and the neglect of higher
order fermion interactions.

The strong coupling stringy NJL model studied here is an example
of a specific NJL-type model, whose virtue is that it can be
easily studied using holography.  Ideally, we would like to be
able to do a corresponding field theory computation of the phase
structure of the same model.  Instead, we just compare against the
simplest large $N$ NJL model, which should therefore be thought of
as a toy model of the real problem.  Also, we will only study the
formation of $\overline{\psi}\psi$ condensates, even though for
real QCD condensates of the form $\psi \psi$ are relevant for
color superconductivity.  Our reason is simply that in our
supergravity description we only saw the appearance of
$\overline{\psi}\psi$ condensates.  A supergravity model of color
superconductivity would of course be interesting to study.

We now turn to a  review of the finite temperature/density phase
diagram of the large $N$ NJL model.   This model can be studied
using standard large $N$ techniques.  Some relevant references are
\cite{AsakawaBQ,Berges:1998rc,SchwarzDJ,Buballa:2003qv}.

The general NJL model is a theory of Dirac fermions $\psi^{ia}$,
with $i=1\ldots N_f$ a flavor index, and $a=1\ldots N_c$ a color
index, that exhibits spontaneous breaking of a continuous chiral
flavor symmetry.  A particular example is
\begin{equation} {\cal L} = \overline{\psi} (i\partial\!\!\!\! / -m)\psi +G
\left[ (\overline{\psi} T^A \psi)^2+(\overline{\psi} i \gamma_5
T^A \psi)^2 \right]~, \label{ha}
\end{equation}
where $T^A$ are generators of $U(N_f)$.  Roughly speaking, we
think of the four-fermi interactions as being induced by single
gluon exchange between quarks.  Of course, in real QCD there is no
justification for omitting multiple gluon exchanges, which
correspond to higher order fermion interactions.     A related
point is that  the four-fermi terms in (\ref{ha}) render this
theory nonrenormalizable, and so a UV cutoff $\Lambda$ needs to be
introduced to define the theory.   While the precise quantitative
predictions derived from this model are strongly cutoff dependent,
the NJL model seems to accurately model the qualitative effects of
chiral symmetry breaking in QCD.

For sufficiently large $G$ this theory develops a nonzero fermion
bilinear condensate.    The mass term $m$ represents explicit
breaking of chiral symmetry, and yields a preferred orientation of
the condensate, namely $\langle \overline{\psi} \psi \rangle \neq
0$, leaving an unbroken diagonal $U(N_f)$ flavor symmetry.   We
then think of taking $m\rightarrow 0$ to remove the explicit
breaking; the motivation for this in the present context is that
the corresponding mass term is absent in the brane picture.

At large $N=N_f N_c$ the standard approach is to introduce an
auxiliary field for the condensing bilinear,
\begin{equation} {\cal L} =\overline{\psi} (i\partial\!\!\!\! /+2 \sqrt{G} \phi)\psi
-\phi^2 +G\left[ (\overline{\psi} T'^A \psi)^2+(\overline{\psi} i
\gamma_5 T^A \psi)^2 \right]~. \label{hb} \end{equation}
Here the notation $T'^{A}$ indicates that the $U(1)$ generator is
omitted. Upon integrating out the auxiliary field $\phi$ we
reproduce (\ref{ha}).

If we instead integrate out the fermions we induce an effective
potential for $\phi$.    The expectation value of $\phi$ yields
the fermion condensate via $\langle \overline{\psi} \psi \rangle =
{1 \over \sqrt{G}} \langle \phi \rangle$.  The leading large $N$
contribution to $V(\phi)$ comes from 1-loop diagrams with a
fermion in the loop and an arbitrary number of external $\phi$
insertions.  From the structure of (\ref{hb}) we see that at this
order we only need the action
\begin{equation} {\cal {L}} =\overline{\psi}
 (i\partial\!\!\!\! /+2 \sqrt{G} \phi)\psi
-\phi^2~. \label{hc} \end{equation}
The effects of the omitted terms are subleading in $1/N$.

\subsection{Zero temperature and chemical potential}

Integrating out the fermion in (\ref{hc}) in the presence of
constant $\phi$ yields the effective potential
\begin{equation} V(\phi) = \phi^2  + i{\rm Tr} \ln ( p\!\!\!/ +M )~, \label{hd}
 \end{equation}
where the effective fermion mass $M$ is
\begin{equation} M= -2 \sqrt{G} \phi~.\label{he} \end{equation}
The ``gap equation" is obtained by minimizing with respect to
$\phi$.  This gives
\begin{equation}  M = 8iN_f N_c G  \int\! {d^4 p \over (2\pi)^4}{M \over p^2
-M^2+i\epsilon~ }~.\label{hda} \end{equation}
Wick rotating and evaluating the momentum integral with a hard
cutoff, $|p| < \Lambda$, we obtain
%
%
%
%
\begin{equation} 1=\lambda
  \left[1 -\left({\Lambda^2 \over M^2}\right)^{-1}\ln\left(1+{\Lambda^2 \over M^2}\right)
  \right]~, \label{hf} \end{equation}
where we defined the dimensionless coupling
\begin{equation} \lambda ={N_f N_c \Lambda^2 G \over 2\pi^2}~. \end{equation}

To determine the condition for chiral symmetry breaking we note
that $f(x)=1-{1 \over x}\ln(1+x)$ is a monotonically increasing
function, obeying $f(0)=0$ and $f(\infty)=1$.  Thus the gap
equation (\ref{hf})\ has no solution for $\lambda <1$.   Since in
going from (\ref{hda}) to (\ref{hf}) we divided by $M$, this
implies that for $\lambda <1$ we have $M=0$, so that chiral
symmetry is unbroken at weak coupling.     For $\lambda >1$ the
gap equation admits a solution with $M\neq 0$, and it is easy to
check that this minimizes $V(\phi)$.    Hence chiral symmetry is
spontaneously broken for $\lambda >1$.

\subsection{Finite  temperature and chemical potential}

We first recall a few basics.  The grand partition sum is
\begin{equation} Z_\Omega = \sum e^{-\beta(E-\mu Q)}~, \end{equation}
and its logarithm gives the grand free energy
\begin{equation}  \Omega(T,\mu) = -{1 \over \beta} \ln Z_\Omega = E - \mu Q -TS~.
\end{equation}
At fixed $T$ and $\mu$ the preferred phase of the system is the
one that minimizes $\Omega$. If we instead work at fixed charge $Q$, then we should  minimize the Helmholtz free energy ,
\begin{equation} F(T,Q) = E-TS =\Omega +\mu Q~. \end{equation}

Returning now to the NJL model, we want to find the gap equation
at finite temperature and chemical potential. It is
straightforward to generalize  the previous derivation (see
\cite{AsakawaBQ,Buballa:2003qv,SchwarzDJ}), but we instead give a
simple physical argument. If we go back to (\ref{hda}) and perform
the $p^0$ integral we get
\begin{equation} M = 4N_f N_c G \int\! {d^3 p \over (2\pi)^3} {M \over E_p}~,
\label{hga}\end{equation}
where $E_p =\sqrt{\vec{p}^2+M^2 }$.   There is an intuitive way to
think about (\ref{hga}): it is the statement that the assumed
effective mass $M$ is indeed the mass obtained by the fermion
interacting with the Dirac sea.  This picture makes it clear  that
to go to finite chemical potential and temperature we instead
write
\begin{equation} M = 4N_f N_c G \int\! {d^3 p \over (2\pi)^3} {M \over
E_p}\Big(1 -n(\beta,\mu)- \overline{n}(\beta,\mu)\Big)~,\label{hia}
\end{equation}
with
\begin{equation} n(\beta,\mu) = {1 \over e^{\beta(E_p -\mu)} +1}~,\quad
\overline{n}(\beta,\mu) = {1 \over e^{\beta(E_p +\mu)} +1}~.
\end{equation}
The thermal gap equation (\ref{hia}) now says that the fermion interacts with the Dirac
sea and the Fermi-Dirac distribution of fermions and anti-fermions
present. This is the same result obtained from a more systematic
derivation.

In the same spirit, the grand free energy is
\begin{equation} \Omega(\beta,\mu) = {M^2 \over 4G} -2N_F N_c \int\! {d^3 p
\over (2\pi)^3} \left\{E_p + {1 \over \beta} \ln \left(1+
e^{-\beta(E_p -\mu)}\right)  + {1 \over \beta} \ln \left(1+
e^{-\beta(E_p +\mu)}\right)\right\}~.\end{equation}
Indeed, extremizing $\Omega$ with respect to $M$ yields
(\ref{hia}).

To map out the phase diagram we need to minimize $\Omega$ at fixed
$(\beta,\mu)$. For future reference the charge density is
\begin{equation} Q(\beta,\mu) = 2N_f N_c\int\! {d^3 p \over (2\pi)^3}
\Big(n(\beta,\mu)- \overline{n}(\beta,\mu)\Big)~.\end{equation}

\subsection{Phase diagram}

In this section we analyze the phase diagram at fixed temperature
and chemical potential.   At low temperatures we find a first
order phase transition that restores chiral symmetry for
sufficiently large chemical potential.   Increasing the
temperature, we eventually find a tricritical point beyond which
the first order transition becomes second order.


We take $G$ sufficiently large such that at zero temperature and
chemical potential there is chiral symmetry breaking (this
corresponds to taking $\lambda >{1 \over 2}$). There is then a
curve $\mu_{\rm crit}(T)$ above which chiral symmetry is restored.
The curve corresponds to the locus of points for which the grand free
energies of the broken and unbroken phases are equal.  At low
temperature, if we raise the chemical potential above $\mu_{\rm
crit}(T)$ we continue to find solutions to the gap equation in the
broken phase.  These solutions are metastable, since the unbroken
phase is thermodynamically preferred for $\mu>\mu_{\rm crit}(T)$.
Continuing to increase the chemical potential, we eventually find
that there no longer exist solutions to the gap equation in the
broken phase, and so the metastable region terminates.   In the
following, we will compute both the actual phase boundary as well
as the metstable region.


Introducing  dimensionless variables via $p = \Lambda \hat{p}~,
\mu = \Lambda \hat{\mu}~, \beta = \Lambda^{-1} \hat{\beta}~, M =
\Lambda \hat{M}$, where $\Lambda$ is the UV cutoff on the
three-momentum, the gap equation becomes
\begin{equation} 1 = 4\lambda \int_0^1 d\hat{p} {\hat{p}^2 \over \sqrt{ \hat{p}^2 +\hat{M}^2}}
\Big(1
-n(\hat{\beta},\hat{\mu})-\overline{n}(\hat{\beta},\hat{\mu})\Big)~.\label{hm}\end{equation}
Here we are using the same symbol for $\lambda$ as in (\ref{hga}),
although the interpretation is slightly different since $\Lambda$
is now a cutoff on 3-momentum rather than 4-momentum. We can also define a rescaled charge density
\begin{equation} \hat{Q}\equiv{\pi Q \over N_f N_c \Lambda^2}= \int_0^1 \!
d\hat{p}~
\hat{p}^2\Big(n(\hat{\beta},\hat{\mu})-\overline{n}(\hat{\beta},\hat{\mu})\Big)~.
\label{hn} \end{equation}

For sufficiently low temperature and chemical potential there
exist solutions to the gap equation (\ref{hm}) with $\hat{M}>0$,
indicating the existence of (stable or metastable) symmetry
breaking vacua. Solutions cease to exist for sufficiently large
temperature and chemical potential, demonstrating symmetry
restoration.  We will first find the boundary between these two
regions.

For definiteness we choose $\lambda =1$.  It is useful to first
consider the special cases of vanishing temperature or chemical
potential.

At vanishing chemical potential (and hence charge) it is easy to
check that the gap equation ceases to have a solution for $\hat{T}
> \hat{T}_c \approx .59$.  Further, as the temperature approaches
$\hat{T}_c$ from below, the equilibrium value of $\hat{M}$ moves
smoothly to the origin.  Thus the phase transition is second order
at finite temperature and vanishing density.

Next consider vanishing temperature and positive chemical
potential.  In this case the distribution functions are
\begin{equation} n(\hat{\beta},\hat{\mu}) = \Theta\left(\hat{p}_F-\hat{p}\right)~,\quad
\overline{n}(\hat{\beta},\hat{\mu}) =0~,\end{equation}
with
\begin{equation} \hat{p}_F=\sqrt{\hat{\mu}^2-\hat{M}^2}~\Theta(\hat{\mu}-\hat{M})~.\end{equation}
 Hence the gap
equation and charge density are
\begin{equation} 1 =4 \int_{\hat{p}_F}^1 {\hat{p}^2 \over
\sqrt{\hat{p}^2+\hat{M}^2}}~,\quad \hat{Q}  = {1 \over 3}
\hat{p}_F^3~.\end{equation}
Consider raising the value of $\hat{\mu}$ from $0$.   At
$\hat{\mu}=0$ we have an uncharged vacuum at $\hat{M}\approx 1.1$.
Clearly, this remains a solution until we reach $\hat{\mu} \approx
1.1$.  As we take $\hat{\mu} >1.1$ we find that there is no
solution to the gap equation, which means that we are driven to
the symmetric vacuum at $\hat{M}=0$. However, well before (at
$\hat{\mu} \approx .83$) we reach this value of $\hat{\mu}$ it is
straightforward to check that the symmetric vacuum has the lower
free energy. Therefore, there is a first order phase transition at
$\hat{\mu} \approx .83$. Note that the transition is between an
uncharged vacuum at $\hat{M} \approx 1.1$ to a charged vacuum at
$\hat{M}=0$.

Finding the complete phase diagram requires a more involved
numerical analysis. We proceed by sampling a number of discrete
choices for $(\hat{T},\hat{\mu},\hat{M})$ and checking whether the
gap equation admits a solution or not.  We then locate the
boundary between the regions with and without solutions.  Given a
solution to the gap equation with $\hat{M}>0$, we need to evaluate
its grand free energy to see if it or the unbroken phase at $\hat{M}=0$
is thermodynamically preferred.  The phase boundary corresponds to
points where the two free energies are equal.  We show the phase
diagram in the $\hat{T}-\hat{\mu}$ plane in figure \ref{figfive}.

\FIGURE[r]{\epsfig{file=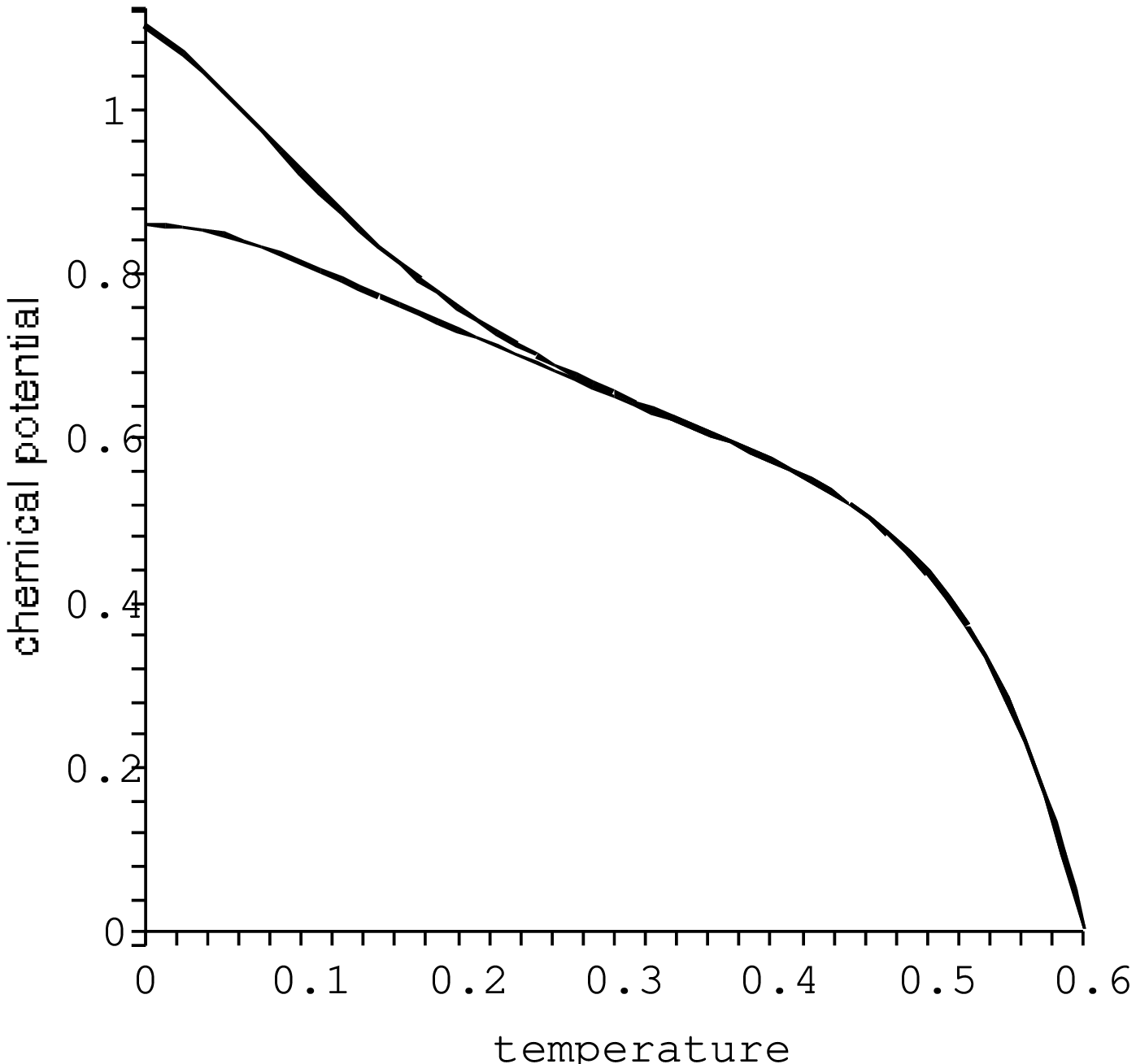, width=6.5cm}
        \caption{  Phase diagram in the $\hat{T}-\hat{\mu}$ plane   }\label{figfive}}

 %
%
The upper curve corresponds to the boundary of the region where
solutions to the gap equation exist.  The lower curve is the true
phase boundary, where the free energies are equal.   In the
temperature regime where the two curves are distinct there is a
first order phase transition, with points between the two curves
representing metastable phases.  This ends at a tricritical point
when the curves meet, and for higher temperatures the phase
transition is second order.

By using (\ref{hn}) we can translate from chemical potential to
charge density.   The phase diagram in the $\hat{T}-\hat{Q}$ plane
is shown in figure \ref{figsix}. There is a tricritical point
where the first order and second order phase boundaries meet.  At
low temperature we have a first order phase transition, and a
corresponding coexistence between the broken and unbroken vacua.
At fixed chemical potential these two vacua have different charge
densities, as indicated in the figure (the smaller of the charge
densities corresponds to the broken phase).

\FIGURE[t]{\epsfig{file=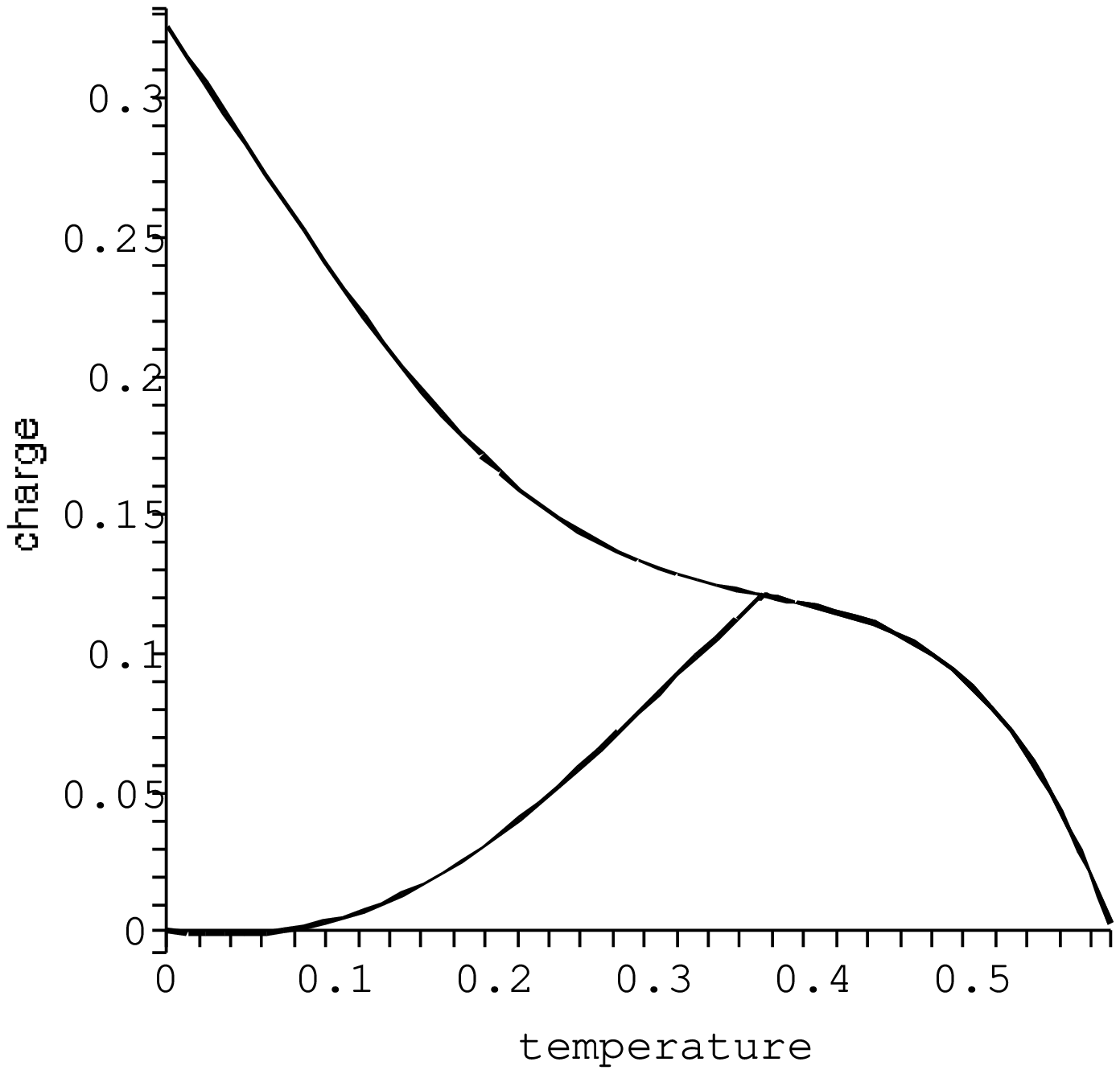, width=10cm}
        \caption{Phase diagram in the $\hat{T}-\hat{Q}$ plane.
        }\label{figsix}}

%
%

For temperatures below the tricritical point, instead of fixing
the chemical potential were we to fix the charge density then the
system will be in a mixed phase, with distinct regions of space
being filled by broken and unbroken phase.  As we increase the
charge density we convert more and more of space into the unbroken
phase.

%
%

At low temperature, the resulting picture gives qualitative
agreement with QCD if we think of the broken phase as vacuum and
the unbroken phase as nuclear matter.  Thus the NJL model can be a
useful guide to studying cold dense matter in QCD.   At higher
temperatures, though, the broken phase of the NJL model starts to
acquire a nonzero charge density, which has no QCD analog. This is
of course due to the lack of confinement, and demonstrates that
the NJL model is a poor guide to high temperature QCD.

\section{Comparison of supergravity versus field theory phase
diagrams}

We can now make a qualitative comparison between the supergravity
and field theory phase diagrams.  It turns out that at low
temperature the phase structures are in reasonable agreement,
while they differ at high temperature.

It is easiest to make the comparison at fixed chemical potential,
which amounts to comparing figures \ref{plotphmu} and
\ref{figfive}. Consider working at fixed temperature and raising
the chemical potential from zero.   At low chemical potential we
are in the broken phase.   As we raise the chemical potential we
eventually enter a range in which the broken phase becomes
metastable.  On the supergravity side this metastable phase  is a
curved brane with no strings attached, hence carrying no charge.
This is in agreement with what we find in the field theory; there
the metastable phase corresponds to $\hat{\mu}<\hat{M}$, which
again implies vanishing charge.   Now raise the chemical potential
further.  On the supergravity side we find that the metastable
curved brane persists but starts to carry a nonzero charge due to
fundamental strings. This aspect differs from the field theory, where for
sufficiently large chemical potential the metastable phase ceases
to exist (there is no solution to the gap equation). On both
sides, the true phase transition is between an uncharged broken
phase and a charged unbroken phase (the straight brane in
supergravity and the $\hat{M}=0$ vacuum in field theory).
Furthermore, the order parameter corresponding to the fermion mass
jumps at the transition, and so the phase transition is first
order in both descriptions.   Altogether, we find that there is
reasonable qualitative agreement between the two sides, including
the existence of metastable phases, with the differences becoming
apparent at the largest chemical potentials.

At higher temperatures the situation is qualitatively different.
The difference is that in field theory there is a tricritical
point at which the phase transition switches from being first
order to second order.   For temperatures above the tricritical
point there is no metastable broken phase. On the supergravity
side the transition is always first order.  A second order
transition would correspond to the branes touching the horizon,
but it turns out that this is never a solution of the equations of
motion.

\section{Stringy Gross-Neveu model}

The 1+1 dimensional analogue of the NJL model is the Gross-Neveu
model \cite{Gross:1974jv}.  In this section we make some
observations about the finite chemical potential thermodynamics of
this theory, in the field theory and brane descriptions.  As we
review, the analysis is essentially trivial since the chemical
potential just couples to a free boson.  Nevertheless, it is
instructive to compare how this comes about in the two
descriptions.

\subsection{Chiral Gross-Neveu model in field theory}

The chiral Gross-Neveu model  is a 1+1 dimensional field theory
with four-fermi interactions invariant under a continuous chiral
symmetry,
\begin{equation} {\cal L} = \overline{\psi}  i\partial\!\!\!\! / \psi +
{\lambda \over 2 N}\left[ (\overline{\psi} \psi)^2 -
(\overline{\psi} \gamma^5 \psi)^2 \right]~.
\end{equation}
Here the fermions are $N$ component objects, with the indices
suppressed and $\lambda$ is the 't Hooft coupling. This theory has
conserved vector and axial currents.

The interaction has the same structure as that induced by
integrating out gluons. Indeed, after using Fierz identities we
can also write the interaction term as
\begin{equation} {\cal L}_{int} ={\lambda \over 2N} ( j^\mu j_\mu + j^{\mu A}
j^A_{\mu})~,\end{equation}
where the singlet and non-singlet currents are
\begin{equation} j^\mu =\overline{\psi} \gamma^\mu \psi~,\quad j^{\mu A} =
\overline{\psi} \gamma^\mu T^A \psi~. \end{equation}
This makes it clear that the theory separates into a $U(1)$ piece
and an $SU(N)$ piece \cite{Affleck}.  The $U(1)$-factor whose
conserved charge is the fermion number (``baryon number") can be
bosonized in terms of a canonically normalized free compact boson
of radius $\sqrt{N \pi}$
\begin{equation} j^\mu =-\sqrt{{N \over 4\pi} }
\epsilon^{\mu\nu}\partial_\nu \phi~, \label{be}\end{equation}
so that fermion number is just the winding number of $\phi$ \cite{Affleck,Schon:2000he}
Adding the interaction just changes the coefficient of the kinetic
term of $\phi$.
%
%
The nontrivial part of the theory is an $SU(N)$ WZW model with
added term $j^A j^A$. Or, equivalently, it is the massless $SU(N)$
Thirring model \cite{Affleck}.

The bosonization (\ref{be}) shows that the free scalar
parameterizes the rotation of the condensate.  We do not actually
have symmetry breaking since massless scalars in two dimensions
cannot sustain expectation values due to IR fluctuations.  But for
large $N$ these fluctuations are suppressed, so we can still think
of the symmetry breaking in an approximate sense.  More precisely,
the relevant two point correlator dies off like $|x|^{-1/N}$
\cite{Witten}.

Adding a chemical potential for baryon number only affects
the $U(1)$ part of the theory, and so is extremely simple to analyze.
In terms of the free boson, a nonzero baryon density is carried by
a linearly varying profile, $\phi = -\sqrt{{4\pi  \over N}} \rho_B
x^1$. Balancing the energy density $\epsilon ={2\pi \over N}
\rho_B^2$ against the chemical potential term $\mu \rho_B$, we
find that the baryon number is proportional to the chemical
potential, $\rho_B ={N\over 4\pi} \mu$. In the fermion language
this just corresponds to filling up the Fermi sea.

The finite temperature thermodynamics is nontrivial since the
$SU(N)$ part of the theory participates.  But again, the effect of
a chemical potential is simple since it only involves the $U(1)$
part. The complete effect of the chemical potential is captured
just by turning on a linearly varying background profile for the
free boson.

%
%

\subsection{Brane version of Gross-Neveu model}

We now consider the non-local generalization  of the Gross-Neveu
model realized as intersecting $D4$ and $D6$-branes
\cite{Antonyan:2006qy}. In the strong coupling limit this system
has an effective description as a $D6-\overline{D6}$ probe
wrapping an $S^4$ in the near horizon geometry of  $N$ $D4$ branes given by (\ref{metricd4}).
The setup is analogous to that giving rise to the stringy NJL
model, the difference being that the dual field theory is now $(1+1)$-dimensional rather than $(3+1)$-dimensional, since the $D8$-branes are
replaced by $D6$-branes.

In the field theory analysis we saw that the theory separates into a $U(1)$
piece and an $SU(N)$ piece.  The $U(1)$ piece is described by a
free boson, which carries the baryon number of the theory via its
winding number.  We would now like to see how this feature emerge
in the strong coupling description.

There are 2+1 unwrapped dimensions of the $D6-\overline{D6}$
probe.  The main ingredient for understanding the above question
is the presence of a Chern-Simons term for the worldvolume gauge
field,
\begin{equation}  S_{CS}= {N \over 4\pi} \int_{M}\! A \wedge
dA~.\label{ea} \end{equation}
where the integration is over the $(1+1)$-dimensional intersection as well as either the $U$ or $x_4$ directions.
To derive this, we start from the RR-coupling on the D6-brane
worldvolume,
\begin{equation} S_6 = {1 \over 2}(2\pi \alpha')^2  T_6 \int\! A \wedge dA
\wedge G_4~. \end{equation}
In the presence of $N$ D4-branes we have $\int_{S^4} \! G_4 =
2\kappa_{10}^2 T_4 N$.  Integrating over the $S^4$ and using $T_4
T_6=[(2\pi)(2\kappa_{10}^2)(2\pi \alpha')^2]^{-1} $ we find the
result (\ref{ea}).

Being the lowest derivative term for the gauge field on the
$D6$-branes, the Chern-Simons term dominates at long distance.  In
the usual spirit of holography we can relate the asymptotic
behavior of the bulk gauge field to the currents and external
gauge fields in the boundary field theory.   To make this precise
we first need to supplement the action (\ref{ea}) with a boundary
term \cite{Elitzur:1989nr,Kraus:2006nb}
\begin{equation} S_{\rm bndy} = -{N\over 8\pi} \int_{\partial M}\!
\sqrt{g}g^{\mu\nu}A_\mu A_\nu~.\label{bndy} \end{equation}
%
Recall that the brane worldvolume boundary  has two distinct
components; denote by $A^{(1,2)}$ the value of the gauge field on
the respective boundary component. With the above choice of
boundary term the on-shell variation of the action is
\begin{equation}\label{bzea}
 \delta S = -{N\over 2\pi} \int\!d^2 x
(A_+^{(1)}\delta A_-^{(1)} + A_-^{(2)}\delta
A_+^{(2)})~,\end{equation}
where $x^\pm$ are lightcone coordinates along the boundary.  From
this expression we can read off the currents
\begin{equation} j_+ = -{N\over 4\pi} A_+^{(1)}~,\quad j_- = -{N\over 4\pi}
A_-^{(2)}~.\label{cur} \end{equation}
Note that the left and right moving currents are each localized on
one of the boundary components. This corresponds to the fact
that the fermions localized at each intersection are chiral (and
have opposite relative chirality). The preceding formulae hold
whether we are in the unbroken phase (straight branes) or broken
phase (curved brane).

Next, we discuss the interpretation of gauge transformations on
the $D6$-branes.   In the unbroken phase we can choose the gauge
$A_U=0$ on the two straight branes, which still leaves us with the
freedom to perform $U$ independent gauge transformation on the
brane and anti-brane, corresponding to an unbroken $U(1)_V \times
U(1)_A$ chiral symmetry.    In the broken phase we can choose
$A_4=0$ along the curved brane.  We then have a single function's
worth of $x^4$ independent gauge transformations, which we
identify with an unbroken $U(1)_V$ symmetry.

The spontaneously broken $U(1)_A$ transformations can be defined
as the gauge transformations
\begin{equation}   \delta A = d\Lambda_A~,\quad \Lambda_A =
f(x_4)\lambda_A(x^+,x^-)~,\end{equation}
where $f(\pm L/2) = \pm 1$.   These symmetries are spontaneously
broken in the sense that they fail to  preserve $A_4=0$.

We expect $U(1)_A$ to act as a shift of the associated Goldstone
boson $\phi$ (the ``pion").    Indeed, if we define $\phi$ as
$e^{i\sqrt{{4\pi\over N}}\phi} = e^{{i\over 2} \int A \cdot dl}$,
where the integration contour connects the two boundary components
at a given $x^\pm$, we see that $U(1)_A$ acts as $\delta \phi =
\sqrt{{N\over 4\pi}}\lambda_A$.   Furthermore, $\phi$ is invariant
under $U(1)_V$, as it should be.  This identification of the
Goldstone boson agrees with that in \cite{Sakai:2004cn}.

The pion $\phi$ is identified with the bosonizing field appearing
in (\ref{be}).    To see this, note that upon acting with $U(1)_A$
we have the currents
\begin{equation}\label{bcea}
 j_+ =-\sqrt{{N\over 4\pi}}\partial_+ \phi~,
\quad j_- =\sqrt{{N\over 4\pi}} \partial_- \phi~,
\end{equation}
which agree with (\ref{be}).

As in the field theory, baryon number is carried by the winding
number of the pion.\footnote{This is a simpler version of the
statement that in the Skyrme model of $3+1$ dimensional $QCD$ a
baryon can be identified with a winding  configuration of the
Goldstone bosons.  In that case, there is a dual description in
terms of an instanton on the brane
\cite{Hata:2007mb,Hong:2007kx}.}   In the brane setup we can turn
on baryon number by performing a large $U(1)_A$ transformation.
The choice $\lambda_A = \sqrt{{4\pi \over N}} \rho_B$ induces the
baryon number density $\rho_B$.  The energy cost for doing this is
entirely governed by the boundary term (\ref{bndy}), since this is
the only nonvanishing contribution to the action in the presence
of a flat connection. The stress tensor following from this
boundary term is that of a free boson, in agreement with the field
theory.

It is now easy to check that the effect of a chemical potential in
the brane setup will agree with that in the field theory.  A
chemical potential $\mu$ is mapped to the boundary conditions
$A_-^{(1)} = A_+^{(2)}= -{1 \over 2} \mu$.  The easiest way to
satisfy this is to just take the constant gauge potential $A_- =
A_+ = -{1 \over 2}\mu$.   Then from ({\ref{cur}) we find that
$\rho_B = j_+ + j_- = {N\over 4\pi} \mu$, which agrees with what
we found in the field theory.

To summarize the results of this subsection, we find that the results in field theory have a simple analogue in the brane
setup.  In field theory, bosonization reduced the problem of
finite baryon density to a problem involving a free scalar field.
On the brane side, everything reduces to pure gauge
configurations. Because the branes have a boundary, these pure
gauge configurations describe physical degrees of freedom.  In
particular, we saw that these degrees of freedom precisely match
those of the bosonizing field.

\section{Landau potentials}

In our supergravity analysis of the phase structure of the stringy NJL and GN models we have
looked for solutions in which there is a smeared distribution of fundamental strings attached
to the probe branes.
A point that we have not emphasized so far is that for generic values of the fundamental string charge  the homogenous solution for the curved brane profile, and therefore the value of the order parameter for the chiral symmetry breaking, is not unique. In field theory near the critical point this type of behavior is typical in the Landau theory of phase transitions.  In this section we address the question of how a Landau potential can be obtained in the supergravity description. At first sight one might conclude that this information is not available on the supergravity side since the correspondence between supergravity  and field theory is on-shell in the bulk.\footnote{See, however, \cite{Sen} for an example where agreement between a certain function evaluated in the CFT and a family of AdS - black holes is found off-shell.} In
general, it is easier to take open strings off-shell than closed strings, since the latter include gravity.
In the present case,  we can certainly take the profile of the probe brane off-shell, while continuing to solve the closed string equations of motion,  and the free energy of this configuration should have a field theory interpretation.

The strategy which we will follow is to take the brane profile
off-shell by giving up the jump condition for the profile function $U(r)$ at the tip of the brane.
As we will see this leads to an effective potential for the order parameter $U_0$ which interpolates between different solutions.    This gives a convenient picture of the phase structure, and
furthermore allows one to assess the (in)stability of the various extrema.    In the following we
will just be considering the zero temperature case for simplicity, but the method is straightforward
to extend to finite temperature.

\subsection{Gross-Neveu  model}
Let us start with the stringy version of the Gross-Neveu model at finite density. The Landau potential in this theory is somewhat artificial since there is no phase transition here as a function of baryon density as explained in \cite{Schon:2000he,Thies}. On the other hand, we can think of studying the phase structure in terms of the fundamental string charge at the tip
of the brane, which in the field theory corresponds to the density of massive fermions.   In this way
we can make contact with the ``old fashioned'' phase diagram of the GN model \cite{Ma}.
In this subsection we show how the corresponding effective potential can be obtained in the gravitational description.

The $D6$ (or $\overline{D6}$) probe brane action in the $D4$-background leads (after integration over the $S^4$) to the 3-dimensional action
\begin{eqnarray}\label{cb}
S_6 &=\sigma\int\! d^3 \xi\, U
e^{-\phi}\sqrt{-\det(g_{ab} + F_{ab}) } \pm  k  \int\!
A\wedge F~,
\end{eqnarray}
where  $\sigma= {T_6 R^3 \Omega_4 \over g_s}$ and $k=\frac{N}{4\pi}$. The metric of the $D4$ background at zero temperature is obtained from (\ref{metricd4}) with $f(U)\equiv 1$. §
For a straight brane, representing the chirally symmetric phase, it is convenient to choose the coordinates $\xi^0=x^0, \xi^1=x^1$ and $\xi^2=U$, whereas for the curved brane we choose $\xi^2=x^4$. It is not hard to show that on the straight brane the only regular solutions with finite action are pure gauge configurations.

In the broken phase the probe brane action reads
\begin{equation}\label{do}
S_6 = \sigma\int\!   d^3\xi ~U \sqrt{
U'^2+ \left({U\over R}\right)^{3} -4 A_+' A_-' ~}~ -~  k
\int\! d^3\xi\,( A_- A_+' - A_+ A_-')~,
\end{equation}
where $'$ denotes differentiation with respect to $x^4$.
The  conserved ($x^4$ independent) quantities are
\begin{eqnarray}\label{df}
 P_\pm &= & {\sigma P_4 \over U^3} {A_\pm'  } \mp k
A_\pm~, \cr
 P_4 & = & {U^4 \over\sqrt{ U'^2+
\left({U\over R}\right)^{3} -4 A_+' A_-' } }~.
\end{eqnarray}
In the absence of sources the only solutions with finite action are again the pure gauge configurations discussed in the last section. We will now consider the curved brane with attached F-strings stretching along $x^4=0$ from the brane tip at $U=U_0$ to the singularity at $U=0$. We will also assume that the F-string is smeared along $x^1$. In field theory this corresponds to switching on a non-vanishing expectation value for the massive fermion in the broken phase.

We construct this solution by taking the two different branches of source-free solutions from above, and matching them across the location of the source.  Then, for a brane
interpolating between $x^4 = \pm \frac{1}{2} L$, we take
\begin{eqnarray}\label{dm}
 A_+ & = \left\{\begin{array}{ccc}
  &  j
-j\exp\left({-{ k \over \sigma P_4}g(U)}\right)~,& \quad x^4<0~,
     \\
  &0~,   &  \quad x^4>0~,       \\
\end{array}\right. \cr
  A_- & =\left\{\begin{array}{ccc}
   & 0~,&  \quad x^4<0~, \\
&  j-j\exp\left({-{ k \over \sigma P_4}g(U)}\right)~,& \quad x^4>0~,
  \end{array}\right.
\end{eqnarray}
where
\begin{equation}\label{dj}
g(U) = \int_{U_0}^U \! dy {y^3 \over \sqrt{{ y^8\over
P_4^2} - \left({y\over R}\right)^3} }~,
\end{equation}
and $U_0$ is expressed through $L$ by
\begin{equation}
{L\over 2} = \int_{U_0}^\infty \! {dy \over \sqrt{{
y^8\over P_4^2} - \left({y\over R}\right)^3} }~.
\end{equation}
The free parameters $j$ and $P_4$ are in turn fixed by matching this solution to the fundamental string source. To this end we add to the brane action
\begin{equation}\label{sf}
S_{F} = {N_1 \over  2\pi  \alpha'}\int\limits_0^{U_0} \!dx^0dU\sqrt{-\det h} - {N_1
\over  2\pi \alpha'}\int\! dx^0 A_0~.
\end{equation}
Requiring the action to be stationary then yields
\begin{equation}\label{c1}
2 V_1  k j={N_1 \over
2\pi  \alpha'}~,
\end{equation}
and
\begin{equation}\label{c2}
P_4^2  =  R^3 U_0^5-{U_0^3 R^3 \over  \sigma^2}\rho^2~,
\end{equation}
where $\rho\equiv {N_1\over{4\pi\alpha'\sigma V_1}}$.
Note that the above solution has no vector $U(1)$-charge. Indeed using (\ref{bzea}) and (\ref{bcea})  we we find $\mu=j$ and $\rho=0$. In field theory language this means that the charge of the massive fermions has been compensated by a negative winding charge, $kj$, of the Goldstone boson $\phi$. From our discussion in section 7.2. it is then clear that the asymptotic charge of this configuration can be adjusted to any value by a suitable axial gauge transformation.\footnote{This gauge choice is actually the right one to compare the energy of this configuration with the straight branes with no gauge field on them.}

For $\rho<\rho_{max}\simeq 0.26\left({R^3 \over L^2}\right)$ the order parameter $U_0$ (which determines the fermion mass through (\ref{sf})) is then a solution of the transcendental equation
\begin{equation}\label{efa}
{1\over 2}{L\over R^{3/2}} = {1 \over \sqrt{\rho} }f(\rho/U_0)~,
\end{equation}
where
\begin{equation}
f(x) = \sqrt{x}\int_1^\infty \!  {dz \over
\sqrt{(1-x^2)^{-1}z^8-z^3}}~.
\end{equation}
For generic  density there are two solutions for the brane profile (see Fig \ref{f1}).
These should correspond to the extrema of the Landau potential for $U_0$. To construct this potential we take the brane profile
off-shell by giving up the jump condition (\ref{c2}) but keeping the condition (\ref{c1}) for  the discontinuity of the gauge fields. We can think of this as displacing the tip of the brane ``by hand''.  To this end we write
\begin{equation}
P_4^2  =  R^3 U_0^5\left(1-x^2\right)~,
\end{equation}
where $x\in [0,1]$ is our off-shell parameter. In this way all equations of motions will be satisfied apart from the jumping
condition for $U'$. On-shell, from the $D6$-brane point of view,
the strength of the discontinuity in $U'$ is a function of the strength
of the external force, in our case, the fundamental string attached to it.\footnote{ This is the analog of the external magnetic field for ferromagnetism.}
$U_0$ is then the ``order parameter'' whose expectation value at
fixed $\rho$ (expressed in units of ${R^3\over L^2}$) is obtained by extremizing
the effective potential $V(U_0)$. It turns out to be easier to determine the potential for $x$ which is related to $U_0$ by integrating the second conservation law in (\ref{df})
\begin{eqnarray}
U_0^{1/2}(x)&=&{{2R^{3/2}}\over{L}}\sqrt{1-x^2}\int\limits_1^\infty{{d z}\over{\sqrt{z^8-(1-x^2)z^3}}}~,\cr
&=&{2R^{3/2}\over 5 L} (1-x^2)^{-1/10}B_{1-x^2}({3\over 5},{1\over 2})~.
\end{eqnarray}
Extremizing $V(x)$ with respect to  $x$ should then imply the jumping condition (\ref{c2}), i.e.
$x=\rho/U_0(x)$.

To continue, we consider the difference in energy between a curved brane with N smeared out fundamental strings and brane profile parameterized by $x$, and two straight branes with no gauge fields on them.
Substitution of the gauge potentials and brane profile into (\ref{cb}), (\ref{sf}) and (\ref{do}) leads to the following expression for the energy difference:
\begin{eqnarray}\label{Le}
{\Delta F(\rho,x)\over 2\sigma V_1}&=&\rho U_0(x)+ U_0(x)^2 \int\limits_1^\infty dz\left({{z^5}\over{\sqrt{z^8-(1-x^2)z^3}}}-z\right)-U_0(x)^2 \int\limits_0^1 dz z~,\cr
&=&\rho U_0(x)+ U_0^2(x)\left\{{(1-x^2)^{2\over 5}\over 5}\int\limits_0^{1-x^2} ds s^{-7\over 5}\left((1-s)^{-1/2}-1\right)-{1\over 2}\right\}~.
\end{eqnarray}
The first term on the r.h.s. of (\ref{Le}) is the tension of the fundamental string (\ref{do}).
The remaining integrals in (\ref{Le}) are regularized incomplete beta functions.
\FIGURE[r]{\epsfig{file=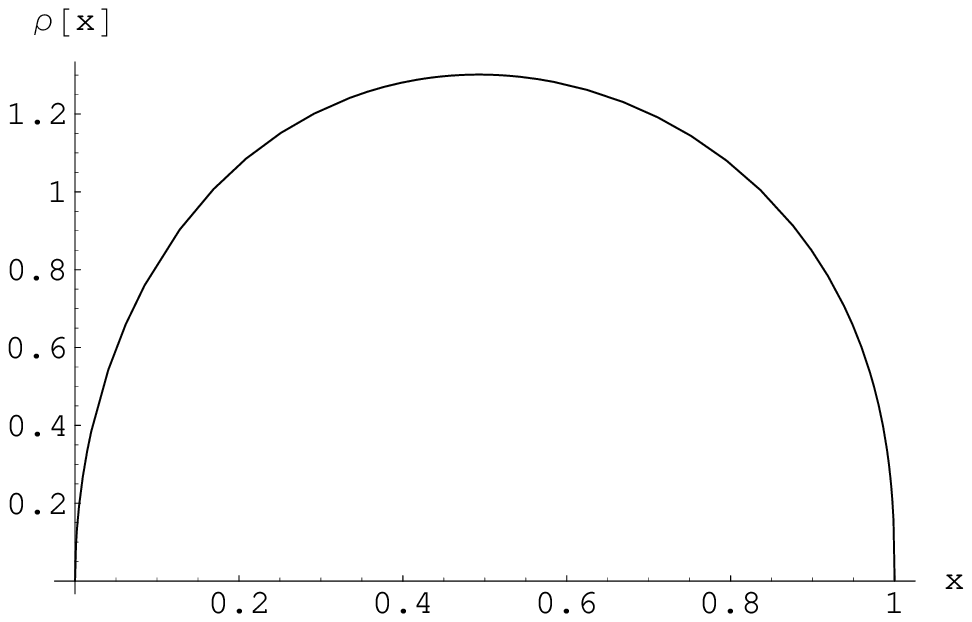, width=7cm}
        \caption{Density, $\rho$ as a function of $x$.}\label{f1}}

The resulting potential is depicted in figure \ref{f2}. The values of $x$ for which  $F(\rho,x)$ attains its local minimum and maximum respectively correspond to the two solutions for $U_0$ of  (\ref{efa}). We will denote the former by $x_{min}$ and the latter by $x_{max}$. We then see that the ``second'' solution (with larger value of $x$) has always higher energy and is therefore unstable.

For $\rho<\rho_c\simeq 0.2$ (in units of ${R^3\over L^2}$) the homogenous curved brane with fundamental strings attached to it has lower energy than the flat brane. Translating this back into field theory units, this gives

\begin{equation}
(N_1/V_1)^{c} \approx   10^{-3} {\lambda \over
2\pi} N\Lambda~.
\end{equation}
For $\rho_c<\rho<\rho_{max}\simeq 0.26$ the curved brane has higher energy than the flat brane (with no density on it). For $\rho>\rho_{max}$ there is no solution for the curved brane.
The vanishing of the Landau potential at $x=1$ may come as a surprise since for non-vanishing density the graphical representation (figure \ref{f1}) of (\ref{efa}) does not give a solution with $x=1$.

\FIGURE[r]{\epsfig{file=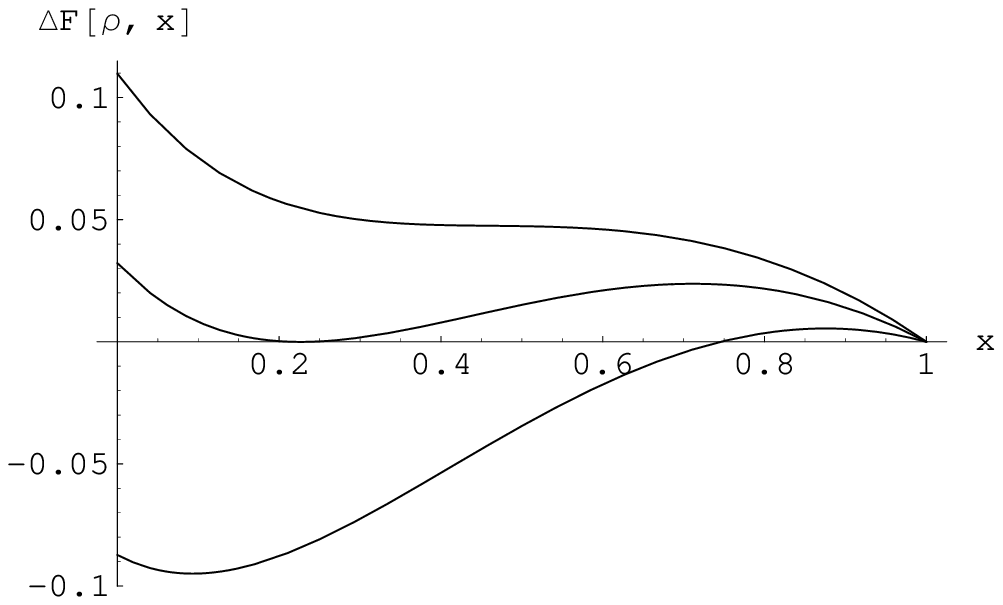, width=7cm}
        \caption{Effective potential for $x$ for $\rho=0.1$, $\rho=\rho_c=0.2$ and $\rho=\rho_{max}=.26$ (in units of $R^3/L^2$).}\label{f2}}

However, inspection of the gauge field configuration (\ref{dm}) shows that since $P_4=0$ for $x=1$, the solutions for the gauge fields on the curved brane have vanishing support so that this configuration is in fact indistinguishable from that of a curved brane with vanishing fundamental string charge and $U_0$, which is in turn a solution of  (\ref{efa}). In other words pulling the tip of charged brane all the way down to the horizon one recovers the uncharged straight brane.

The picture that arises here appears to be in qualitative agreement with the mean field approximation (see \textit{e.g.} \cite{Thies})
assuming one assigns a unit charge to the fundamental string. Then our effective potential for the mean field $U_0$ predicts a phase transition to the chirally symmetric phase at $\rho=\rho_c$ in agreement with the (old fashioned) phase diagram for the GN model.

Finally one can obtain the mass $M$ of the ``mean field fermion'' as
\begin{eqnarray}
{2 \pi\alpha' M}&\equiv&{\partial E\over \partial \rho}|_{\rho=0}~,\cr
&=&{3\over 5}U_0+{2\over 5}{\partial U_0\over \partial x}|_{x=0}
\left\{\int\limits_0^{1} ds s^{-7\over 5}\left((1-s)^{-1/2}-1\right)-B({2\over 5},1)\right\}~,\cr
&=&{8\over 5}U_0~.
\end{eqnarray}
This is not the physical fermion mass in general. Rather it is the mass of a delocalized fermion obtained by assuming that single particle states are momentum eigenstates (which would be correct for a free theory). If these massive fermions were non-interacting we should get $2 \pi\alpha' M=U_0$.

\subsection{Stringy NJL model}
We now turn to the stringy NJL model. Again we will consider the model at finite density but zero temperature for simplicity. In this case the matching conditions (\ref{jumpa}) for a smeared fundamental string source become\footnote{We absorb $\beta$ into the definition of the charge.}
\begin{equation}\label{ssbt}
Q={N_1\over 2 \pi \alpha'}\equiv \rho\kappa R^{\frac{3}{2}}~,
\end{equation}
and
\begin{equation}\label{pfo}
P_4^2=U_0^8\left(\kappa^2-{\rho^2\kappa^2\over U_0^5}\right)+Q^2\left(\frac{U_0}{R}\right)^3~.
\end{equation}
Upon substitution of (\ref{ssbt}) the later leads to (\ref{jumpb}) at $T=0$.
We could now define the off-shell Landau potential by relaxing the constraint  (\ref{jumpb}) for $P_4$ while keeping (\ref{ssbt}) as we did in the case for the $D6$-brane. However, unlike for the $D6$-brane where $Q$ did not appear in the DBI action and the expression for $\dot U$,  this ansatz
leads to complicated expressions in which $U_0$ is only implicitly determined, so that one has to solve the corresponding system numerically as in section 4. For the sake of computational simplicity we will therefore follow a different strategy, in which both (\ref{ssbt}) and (\ref{pfo}) are taken off-shell simultaneously by introducing a single parameter $x$. Since both  (\ref{ssbt}) and (\ref{pfo}) are
equations of motions for the combined system of probe brane and fundamental strings,
this should be a consistent off-shell definition, at least near the critical points. Concretely we set
\begin{equation}\label{sabtt}
x^2={Q^2\over \kappa^2 R^3 U_0^5(x)}~,\qquad\hbox{and}\qquad P_4(x)=\kappa U_0^4(x)~,
\end{equation}
where  $U_0(x)$ is determined by integrating (3.13), \textit{i.e.}
\begin{equation}
U_0(x)={4R^{3}\over L^2} (g(x))^2~,
\end{equation}
with
\begin{equation}
g(x)=\int\limits_1^\infty {d z\over\sqrt{z^{11}-z^3+z^6 x^2}}~.
\end{equation}
Note that if (\ref{ssbt}) is not satisfied then $P_4$ as defined in (\ref{pfo}) is taken off-shell as well, since the charge no longer drops out in (\ref{pfo}). For a generic value of $x$ the jumping conditions are satisfied for some charge density which is different from that sourced by the fundamental string charge $\rho$. To see this we plot
\begin{equation}\label{ssmax}
{L( \rho)^{1\over 5}\over 2 R^{3\over 2}}= x^{1\over 5} g(x)~,
\end{equation}
against $x$ (see figure \ref{f3}).  For $\rho < \rho_{max}\simeq  (0.47)^5({R^{3\over2}/ L})^5$ there are two solutions for fixed fundamental string charge $\rho$. To see which solution has lower energy we compute the Landau free energy off-shell as a function of $x$ at fixed $\rho$. The actual solution is obtained by minimizing the Landau free energy with respect to $x$.
In order to get the free energy we take the Legendre transform of the brane and string action, as explained in section 4,
\begin{equation}\label{fe}
F(Q)=2\int\limits_{U_0}^\infty L_{DBI}(A,A')+S_F+2Q A(\infty)~.
\end{equation}
The factor of 2 takes into account both legs of the curved brane. For a given charge  $Q(x)$ on the curved brane we have
\begin{equation}
A^{curved}(x,\infty)=xU_0(x)\int\limits_{1}^\infty{d z z^{3\over 2}\over\sqrt{z^{8}-1+z^3 x^2}}~,
\end{equation}
while for a straight brane with charge $\rho$ we have
\begin{equation}
A^{straight}(\rho,\infty)=\rho^{2\over 5}{\Gamma({3\over 10})\Gamma({6\over 5})\over \sqrt{\pi}}~.
\end{equation}
The free energy difference between the curved brane with charge $Q(x)$ and the straight brane with charge $\rho$ is then given by\footnote{Since the Legendre transform of the
grand canonical potential is well defined only on-shell (i.e. when $Q$ is the physical charge of the system) there is some ambiguity in defining $F(\rho,x)$. Here we take the point of view where the $Q$ entering in (\ref{fe}) is the physical charge.}
\begin{eqnarray}\label{feza}
\Delta F(\rho,x)&=&-S_{F}(\rho,U_0(x))+2\rho \kappa R^{3\over 2}(A^{c}(x,\infty) -A^{s}(\rho,\infty))\cr
&&~~-2S^{c}_{DBI}( A'(x))+2S^{s}_{DBI}( A'(\rho))~.
\end{eqnarray}
Putting it all together we get
\begin{eqnarray}\label{ssEL}
{\Delta F(\rho,x)\over 2\kappa R^{3\over 2}}&=&\rho U_0(x) +\rho xU_0(x)\int\limits_{1}^\infty{d z z^{3\over 2}\over\sqrt{z^{8}-1+z^3 x^2}}-\rho^{7\over 5}{\Gamma({3\over 10})\Gamma({6\over 5})\over \sqrt{\pi}}\cr
&&+U_0(x)^{7\over 2} \left\{
\int\limits_1^\infty {d z z^8\over\sqrt{z^{11}-z^3+z^6 x^2}} -
\int\limits_0^\infty d z z^{5\over 2}\sqrt{z^5\over z^5+{\rho^2\over U_0^5(x)}}\right\}~,
\end{eqnarray}
where the order of the different terms in (\ref{ssEL}) is the same as in (\ref{feza}).

\FIGURE[t]{\epsfig{file=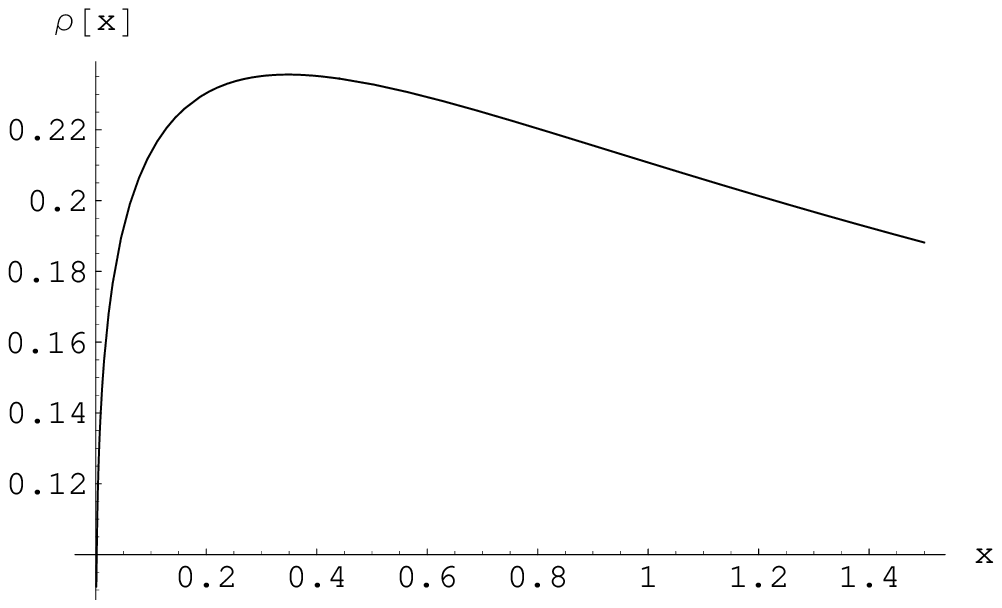, width=10cm}\label{f3}
        \caption{Density $\rho$ as function of $x$ as in (\ref{ssmax}).}}

The remaining integrals are easily evaluated numerically. Figure \ref{f4} showing the effective potential for $x$ reveals a picture that is in quantitative agreement with figure \ref{f3} and (\ref{ssmax}) in particular. For given charge density $\rho$ with $\rho<\rho_{max}$ the effective potential has a local minimum, which
corresponds to the smaller value of the $2$-possible solutions of (\ref{ssmax}), and a
local maximum which corresponds to the larger value of $x$. The actual value of $\rho_{max}$ obtained by analyzing the effective potential is identical with (\ref{ssmax}) (see figure \ref{f4}). Another feature of the effective potential is that the energy of the curved brane with finite charge density exceeds that of the straight branes for $\rho>\rho_c$ with
$\rho_c\simeq (0.23)^5$. The Landau potential predicts a first order transition to the chirally symmetric phase.

\FIGURE[r]{\epsfig{file=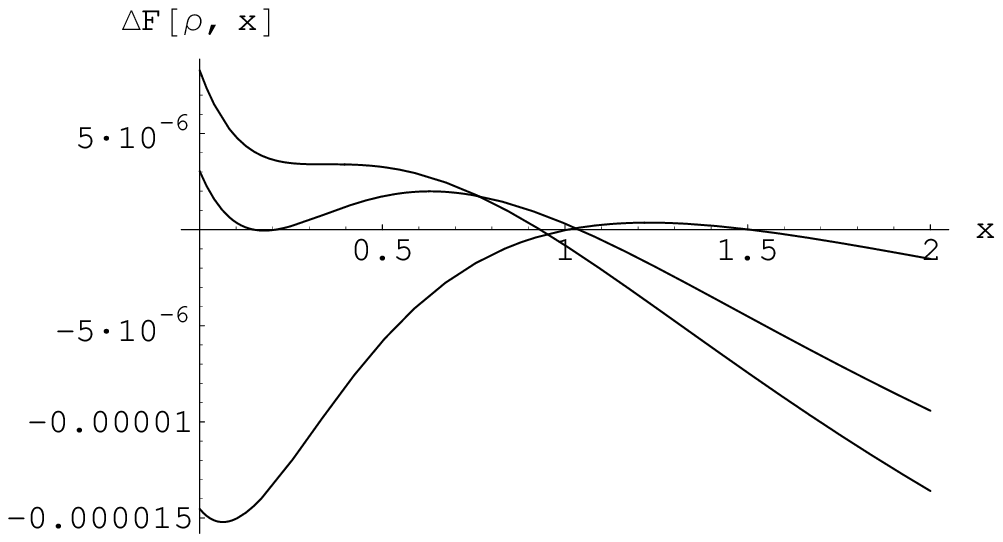, width=7cm}
        \caption{$\Delta F(\rho,x)$ as a function of $x$ for $\rho=(0.2)^5$, $\rho=\rho_c=(0.23)^5$ and $\rho=\rho_{max}=(0.24)^5$ in units of $(2R^{3/2}/L)^5$.}\label{f4}}

As explained in section 5 the mean field approximation discussed here does not yield the correct phase diagram at large $N_c$. The physically realized phase at non-vanishing density is rather a inhomogeneous  mixture of phases of uncharged curved branes and charged straight branes. Although we will not repeat the analysis here, it is not hard to show that the free energy of the mixed phase is lower than the minimum of the Landau free energy of the pure phase. The corresponding phase diagram is identical with that obtained in section 5 for zero temperature.

We conclude this section by noting that the Landau potential does not vanish for $x\to\infty$ unless $\rho=0$. The fall-off of $\Delta F$ for $x\to\infty$ is a consequence of the unphysical relaxation of the matching condition for the charge in (\ref{ssbt}). The corresponding Landau potential is unphysical  for large $x$. In particular, the fall-off at large $x$ should not be interpreted as an instability of the configuration corresponding to the local minimum.

\section{Discussion}

To conclude, we briefly summarize the main results of this paper.   We have mapped out the
finite density/temperature phase structures of the stringy NJL and GN models at strong
coupling by studying probe branes in the near horizon geometry of $D4$-branes.   In the stringy NJL
model an important role was played by solutions with fundamental strings having one end on  the probe branes and the other end disappearing through the $D4$-brane horizon. These fundamental strings carry
baryon number in the phase with broken chiral symmetry.
We also reviewed the corresponding phase diagrams in the local field theory cousins of these models,
and compared with the strong coupling supergravity results.  For the stringy NJL model we saw
good qualitative agreement in the phase structure at low temperatures, except at the very highest
densities.  On the other hand, at high temperatures there is a qualitative difference between the
two sides, the chiral phase transition being respectively first and second order in the supergravity and field
theory models.    For the stringy GN model there is a precise agreement between the two sides,
but in a somewhat trivial way.  In both supergravity and field theory, the charge density is simply  carried
by  a free boson.  In supergravity the free boson corresponds to pure gauge modes on the probe brane,
and in field theory it appears via bosonization.

In our study of the phase structure we encountered various stable, unstable, and metastable phases.
For instance, when we looked for curved brane solutions with attached fundamental strings,
we found that the equations of motion admitted multiple brane profiles with a given charge
density.  A convenient way to keep track of these different solutions is via an off-shell Landau
potential.   By taking the probe branes off-shell in a particular way, we can interpolate between
the various solutions, and the absolute minimum of the Landau potential determines the
stable phase.   We illustrated this procedure in the simplified case of zero temperature.

As discussed in the introduction, the long term motivation for the analysis considered
here is to eventually apply stringy methods to learn about finite density QCD.   Since NJL type
models are one of the main phenomenological tools in this subject, studying their string theory
incarnations is potentially useful.     The main challenge ahead is to find ways to modify the stringy models
 so as to make them more closely resemble physical QCD.

\acknowledgments
We thank Jeff Harvey for correspondence.
The work of JD, MG and PK was supported in part by National Science Foundation (NSF) Physics Division
grant PHY-04-56200. The work of IS was supported in parts by the SPP-1096, the Transregio TR-33 and the Excellence Cluster "Origin and Structure of the Universe" of the DFG.  IS. would like to thank the Department of Physics and Astronomy at UCLA and the Physics Department at Caltech for hospitality during the initial stage of this work.

\end{document}